\documentclass[preprint]{aastex}
\usepackage[]{graphics}
\usepackage[]{psfig}
\usepackage[]{epsfig}
\usepackage[]{longtable}
\usepackage[]{natbib}
\shorttitle{UCDs in Abell 1689 with ACS}
\shortauthors{Mieske et al.}

\begin{document}

\title{Ultra Compact Dwarf galaxies in Abell 1689: \\
a photometric study with the ACS}

\author{S. Mieske \altaffilmark{1}, L. Infante}
\affil{Departamento de Astronom\'{\i}a y Astrof\'{\i}sica, Pontificia
Universidad Cat\'olica de Chile, Casilla 306, Santiago 22, Chile}
\email{smieske@astro.uni-bonn.de,linfante@astro.puc.cl}

\author{
N. Ben\'{\i}tez\altaffilmark{2},
D. Coe\altaffilmark{2},
J.P. Blakeslee\altaffilmark{2},
K. Zekser\altaffilmark{2},
H.C. Ford\altaffilmark{2},
T.J. Broadhurst\altaffilmark{6},
G.D. Illingworth\altaffilmark{4},
G.F. Hartig\altaffilmark{3},
M. Clampin\altaffilmark{8},
D.R. Ardila\altaffilmark{2}
F. Bartko\altaffilmark{5}, 
R.J. Bouwens\altaffilmark{2},
R.A. Brown\altaffilmark{3},
C.J. Burrows\altaffilmark{3},
E.S. Cheng\altaffilmark{7},
N.J.G. Cross\altaffilmark{2},
P.D. Feldman\altaffilmark{2},
M. Franx\altaffilmark{9},
D.A. Golimowski\altaffilmark{2},
T. Goto\altaffilmark{2},
C. Gronwall\altaffilmark{10},
B. Holden\altaffilmark{4},
N. Homeier\altaffilmark{2},
R.A. Kimble\altaffilmark{8},
J.E. Krist\altaffilmark{3},
M.P. Lesser\altaffilmark{11},
A.R. Martel\altaffilmark{2},
F. Menanteau\altaffilmark{2},
G.R. Meurer\altaffilmark{2},
G.K. Miley\altaffilmark{9},
M. Postman\altaffilmark{3},
P. Rosati\altaffilmark{12}, 
M. Sirianni\altaffilmark{3}, 
W.B. Sparks\altaffilmark{3}, 
H.D. Tran\altaffilmark{13}, 
Z.I. Tsvetanov\altaffilmark{2},   
R.L. White\altaffilmark{3}
\& W. Zheng\altaffilmark{2}}


\altaffiltext{1}{Sternwarte der Universit\"at Bonn, Auf dem H\"ugel 71,
  53121 Bonn, Germany}

\altaffiltext{2}{Department of Physics and Astronomy, Johns Hopkins
University, 3400 North Charles Street, Baltimore, MD 21218.}

\altaffiltext{3}{STScI, 3700 San Martin Drive, Baltimore, MD 21218.}

\altaffiltext{4}{UCO/Lick Observatory, University of California, Santa
Cruz, CA 95064.}


\altaffiltext{5}{Bartko Science \& Technology, 14520 Akron Street, 
Brighton, CO 80602.}    


\altaffiltext{6}{Racah Institute of Physics, The Hebrew University,
Jerusalem, Israel 91904.}


\altaffiltext{7}{Conceptual Analytics, LLC, 8209 Woburn Abbey Road, Glenn Dale, MD 20769}

\altaffiltext{8}{NASA Goddard Space Flight Center, Code 681, Greenbelt, MD 20771.}


\altaffiltext{9}{Leiden Observatory, Postbus 9513, 2300 RA Leiden,
Netherlands.}


\altaffiltext{10}{Department of Astronomy and Astrophysics, The
Pennsylvania State University, 525 Davey Lab, University Park, PA
16802.}


\altaffiltext{11}{Steward Observatory, University of Arizona, Tucson,
AZ 85721.}


\altaffiltext{12}{European Southern Observatory,
Karl-Schwarzschild-Strasse 2, D-85748 Garching, Germany.}


\altaffiltext{13}{W. M. Keck Observatory, 65-1120 Mamalahoa Hwy., 
Kamuela, HI 96743}

\begin{abstract}
\noindent The properties of Ultra Compact Dwarf (UCD) galaxy candidates in
Abell 1689 (z=0.183) are investigated, based on deep high resolution
ACS images. A UCD candidate has to be
unresolved, have $i<28$ ($M_V < -11.5$) mag and satisfy color limits
 derived from Bayesian photometric redshifts. 
We find 160 UCD candidates with $22<i<28$ mag.
We estimate that about 100 of these are cluster
members, based on their spatial distribution and
photometric redshifts. For $i \gtrsim 26.8$ mag, the radial and 
luminosity distribution of the UCD candidates can be explained well by
Abell 1689's globular cluster (GC) system. For $i \lesssim 26.8$ mag,
there is an overpopulation of 15 $\pm$ 5 UCD candidates with respect to
the GC luminosity function. For $i \lesssim 26$ mag, the radial distribution
of UCD candidates is more consistent with the dwarf galaxy population than
with the GC system of Abell 1689. The UCD candidates follow a
color-magnitude
trend with a slope similar to that of Abell 1689's genuine dwarf galaxy
population, but shifted fainter by about 2-3 mag. Two of
the three brightest UCD candidates ($M_V\simeq -17$ mag) are slightly
resolved. At the distance of Abell 1689, these two objects would have 
King-profile core radii of $\simeq$ 35 pc and $r_{\rm eff}\simeq$ 300
  pc, implying luminosities and sizes 2-3 times those of M32's bulge. 
Additional photometric redshifts obtained
with late type stellar and elliptical galaxy templates support the
assignment of these two resolved sources to Abell 1689, but also allow for up to 4
foreground stars among the 6 brightest UCD candidates.
Our findings imply that in Abell 1689 there are $\ge$ 10 UCDs with
$M_V<-12.7$ mag, probably created by stripping ``normal'' dwarf or
  spiral galaxies. 
Compared to the UCDs in the Fornax cluster -- the location
of their original discovery -- they are
brighter, larger and have colors closer to normal dwarf galaxies. This suggests
that they may be in an intermediate stage of the stripping process. 
Checking the photometric 
redshifts of the brightest UCD candidates with spectroscopy would be the next
step in order to definitely confirm the existence of UCDs in Abell 1689.
\end{abstract}


\section{Introduction}
\subsection{Discovery of UCDs}
\noindent Recently, Drinkwater et al.~\cite{Drinkw00} and~\cite{Drinkw03}, reported
 on the discovery of 7 ultra compact dwarf galaxies (UCDs) in the Fornax
cluster. These objects are very luminous star clusters in the magnitude range
$-13.4<M_V<-12$ mag, i.e. about 1-2 mag brighter than $\omega$ Centauri. Three
different origins for the UCDs are in discussion: (1) they are the brightest
globular clusters of very rich globular cluster systems (GCS), like in NGC 1399
(Mieske et al. \cite{Mieske02}, Dirsch et al. \cite{Dirsch03}); (2) they
are the remnant nuclei of stripped dwarf galaxies which have lost their outer
parts in the course of tidal interaction with the Fornax cluster's potential
(Bekki et al.~\cite{Bekki03}, Mieske et al.~\cite{Mieske04}); (3) they are
formed from the amalgamation of stellar super-clusters in collisions between
gas-rich galaxies (Fellhauer \& Kroupa~\cite {Fellha02}, Kroupa
\cite{Kroupa98}, Maraston et al.~\cite{Marast04}).

These possibilities are discussed in more detail in Mieske et al.
\cite{Mieske04}. It is found that in Fornax there are 12 bright compact
objects with $-13.4<M_V<-11.4$ mag, including the UCDs. Applying
incompleteness correction, this number rises to 14. The bright compact objects
follow a color-magnitude relation in $(V-I)$ with a slope very similar to
that of Fornax dEs (Hilker et al.~\cite{Hilker03}), but shifted about 0.2 mag
redwards. In addition, they appear to be separated from the fainter compact
Fornax members ($M_V>-11.4$ mag) in radial velocity and are spatially
distributed in a more extended fashion. The properties of the faint compact
Fornax members are consistent with the globular cluster system of NGC 1399.
The properties of bright compact Fornax members seem to be different from those
of normal globular clusters. They are consistent with the threshing scenario
of Bekki et al.~\cite{Bekki03}, who propose tidal stripping of nucleated dwarf
galaxies as a source of bright compact cluster members. Their properties are
also consistent with the super-cluster scenario as proposed by Fellhauer \&
Kroupa~\cite{Fellha02}, suggesting for the first time a color magnitude
relation for bright globular clusters, if true.

If UCDs are bright compact stellar systems distinct from globular clusters, the
results of Mieske et al.~\cite{Mieske04} then show that UCDs in Fornax
populate the color-magnitude range $-13.4<M_V<-11.4$ mag and $1.0<(V-I)<1.30$
mag, extending the discoveries by Drinkwater et al. \cite{Drinkw03} to
fainter limits.
\subsection{UCDs in Abell 1689?}
\noindent In the case of Abell 1689 ($z=0.183$, Tyson \& Fischer
\cite{Tyson95}) 
one of the most massive known galaxy clusters ($M \simeq 0.5-2 \times 10^{15} M_*$,
$M/L \simeq 400$ and $r_s \simeq 350$ kpc, see the lensing studies by
Broadhurst et al.~\cite{Broadh04} and King et al.~\cite{King02}) it is very
interesting to estimate the number of UCD candidates. Is Fornax a special
case, or are UCDs a more general phenomenon? A consequence of the latter
possibility would be that entire tidal disruption of fainter dwarf galaxies
might occur in many clusters and could therefore partially cause the
``missing satellite'' problem (Moore et al.~\cite{Moore99}, Klypin et
al.~\cite{Klypin99}). 
Detailed numerical simulations regarding
this
issue might become necessary. However, it must also be addressed
to what extent the number of
UCD candidates can be accounted for by the very rich GCS of Abell 1689 (Blakeslee
et al.~\cite{Blakes03}). As Abell 1689 consists of several sub-clusters in
radial velocity (Teague et al.\cite{Teague90}, Girardi et al.\cite{Girard97}),
there is the possibility of merger events having occurred in the recent past.
This might cause the creation of stellar super-cluster (Fellhauer \& Kroupa
\cite{Fellha02})and therefore also contribute to the number of UCDs.
Abell 1689 has a mean redshift of z=0.1832 (NED database). Assuming $H_0=70$
km/s/Mpc, $\Omega_M=0.3$ and $\Omega_{\Lambda}=0.7$, the corresponding
distance modulus $(m-M)$ is 39.74 mag. Taking into account k-correction,
which is about 0.3 mag for $V$ at the redshift of Abell 1689, we look for objects in
the apparent magnitude range $26.65<V<28.65$ mag, a task which requires very
deep imaging.

 The aim of this paper is to estimate the number of UCD candidates
 in Abell 1689 and to investigate their distribution in luminosity, color, and
 space using deep ACS imaging data. Sect.~\ref{data} describes the ACS data.
 In Sect.~\ref{ncalc} the properties of UCD candidates are investigated. In
 Sect.~\ref{discussion}, the findings are discussed. Sect.~\ref{conclusions}
 presents the conclusions.

\section{The data}
\label{data}

\noindent The photometric data used in this paper are extracted from deep ACS
images of Abell 1689, obtained from the ACS Guaranteed Time Observations deep
imaging cluster Wide Field Camera (WFC) data in June 2002. The WFC covers a
FOV of 202 $\times$ 202 arcsec, with 0.05$''$ pixels. The data are presented
in more detail in Broadhurst et
al.~\cite{Broadh04}. A total of 20 orbits were
taken in the four passbands F475W(g), F625W(r), F775W(i) and F850L(z),
corresponding to the Sloan (g´,r´,i´,z´) Digital Sky Survey (SDSS) filters.
The PSF full-width at half-maximum (FWHM) is 0.10$''$--0.11$''$,
   or about 2 pixels, in each filter.
The
limiting magnitude for detecting point sources with at least 5 pixel above 
1.5 $\sigma$ is $i\simeq 28.5$ mag. The faint
limit of the UCD magnitude regime of $V=28.65$ mag corresponds to $i\simeq
28.0$ mag (Fukugita et al.~\cite{Fukugi96}).

At the distance of Abell 1689, 0.1$''$=310 pc. All UCDs in 
Fornax have scale
lengths of 10-20 pc. Their analogs hence are detectable but unresolved on the
ACS images of Abell 1689.

Detection and analysis of unresolved sources was done using SExtractor (Bertin
\& Arnouts~\cite{Bertin96}) on the ACS images, where all bright galaxies were
previously  subtracted (see Zekser et al.\cite{Zekser04}). SExtractor was run
in dual mode, detection of sources on the detection image (inverse-variance weighted average
of the four passband images, see Benitez et al. \cite{Benite04}), 
and then, carrying out their analysis on each single passband.
Sources were defined as unresolved when the SExtractor Star-Classifier value
was larger than 0.6. Magnitudes were determined from aperture photometry.

Using the technique outlined by Benitez~\cite{Benite00}, Coe et
al.~\cite{Coe04} provide Bayesian Photometric Redshifts (BPZ) for almost
2000 sources in Abell 1689. The ACS $griz$ data was complemented by VLT
optical and infrared photometry. In this paper these photometric redshifts are
used to help distinguish unresolved cluster members from background galaxies.

The analysis of the resolved sources and thereby of the normal dwarf galaxy
population in Abell 1689 is presented in a separate paper (Infante et
al.~\cite{Infant04}).
 
\section{Photometric selection and properties of UCD candidates in Abell 1689}
\label{ncalc}
In this paper two different methods to separate cluster members from background
galaxies are applied: first, a color selection applied to unresolved objects, based on
and complemented by photometric redshifts; and
second, a statistical background subtraction using the radial density
distribution of the color selected objects. The background contaminations
derived from both methods are compared. To search for UCD candidates, 
a circular region of 92 $''$ (285 kpc) radius centered on the brightest
cluster galaxy was analysed. This was the largest possible circular region
imaged entirely by the ACS. See the paper by Broadhurst et
al.~\cite{Broadh04} for the optical ACS image of the cluster.

\subsection{Color selection and background contamination from photometric redshifts}

\begin{figure}
\plotone{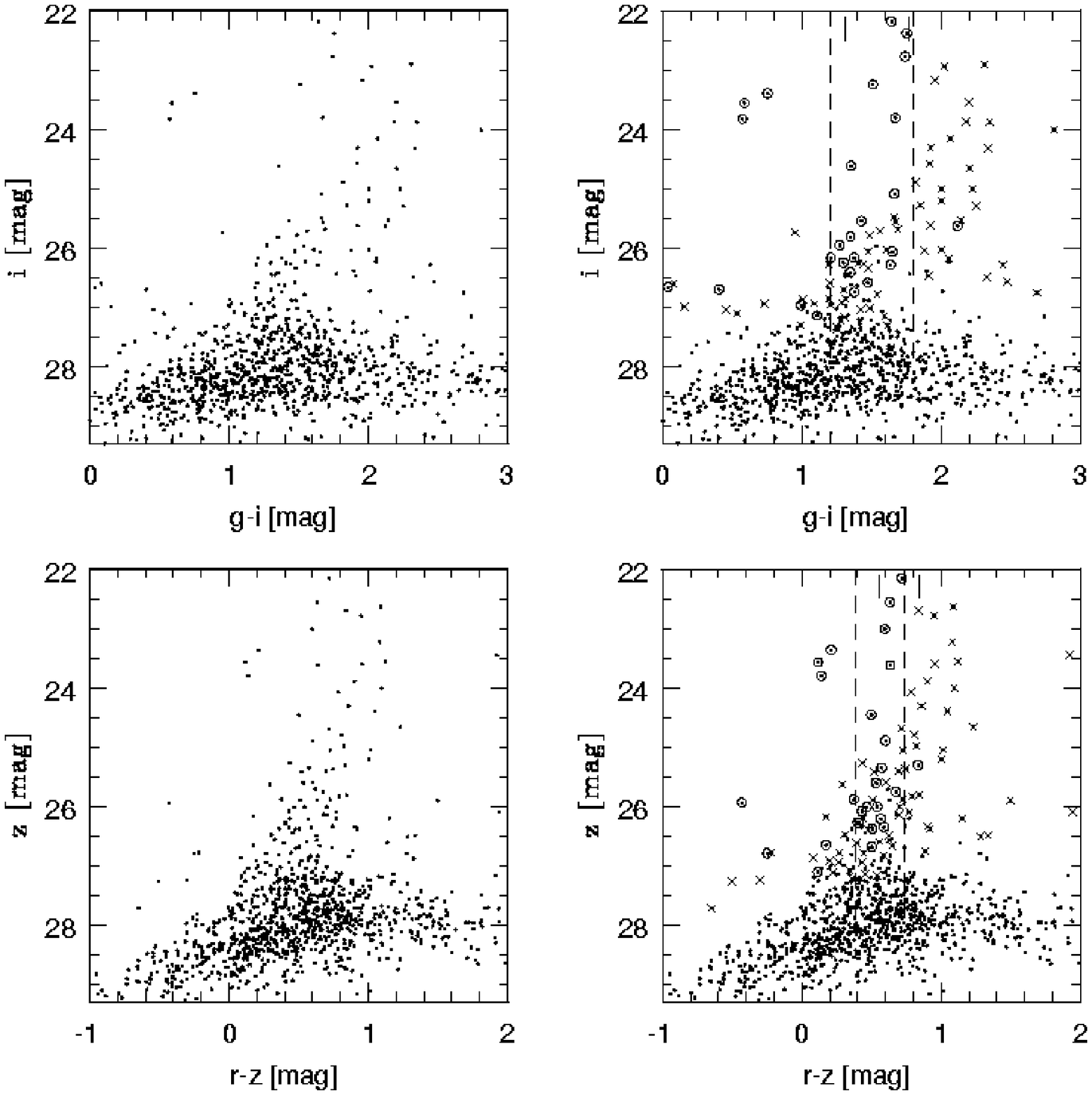}
\caption{Top
panel left: CMD in $i$ and $(g-i)$ of all unresolved sources in the central
92$''$ (285 kpc) of Abell 1689. Top panel right: dots as plotted to the left,
circles indicate objects with photometric redshift $z_{\rm phot}\le 0.5$
(taken from Broadhurst et al.~\cite{Broadh04} and Coe et al.~\cite{Coe04}), crosses objects with $z_{\rm
phot}>0.5$. Vertical ticks at the top indicate the blue and red color limit of
UCDs from Fornax (Mieske et al.~\cite{Mieske04}), transformed from $VI$ to
$gi$ using the calibrations of the ACS bandpasses described in Sirianni et al.~\cite{Sirian04}.
Dashed vertical lines indicate the finally adopted color window in
$(g-i)$ for UCD candidates in Abell 1689, extending slightly more blue-wards
compared
to the transformed UCD colors in Fornax. Bottom panels: Same as top panels,
but in $z$ and $(r-z)$. The color selection window is shifted about 0.15 mag blue-wards
compared to the transformed UCD colors. \label{cmdall}}
\end{figure}

\noindent The left panels of Fig.~\ref{cmdall} show two CMDs in $(g-i)$ and
$(r-z)$, respectively, of all unresolved sources in the ACS field of Abell
1689. The faint magnitude limit for UCD candidates is $i=28$ mag.
To help define color selection windows for UCD candidates, photometric
redshifts $z_{\rm phot}$ from Coe et al.~\cite{Coe04} are indicated in
the right panels of Fig.~\ref{cmdall} to separate cluster member candidates
from background galaxies: Circles are for $z_{\rm phot}\le0.5$, and crosses
are for $z_{\rm phot}>0.5$. In addition, the color windows in $(g-i)$ and
$(r-z)$ corresponding to the $VI$ colors of UCDs in Fornax are also indicated.
The color transformations were calculated using the calibrations 
of the ACS bandpasses described in Sirianni et al.~\cite{Sirian04},
including $k$-corrections.

\begin{figure}
\plotone{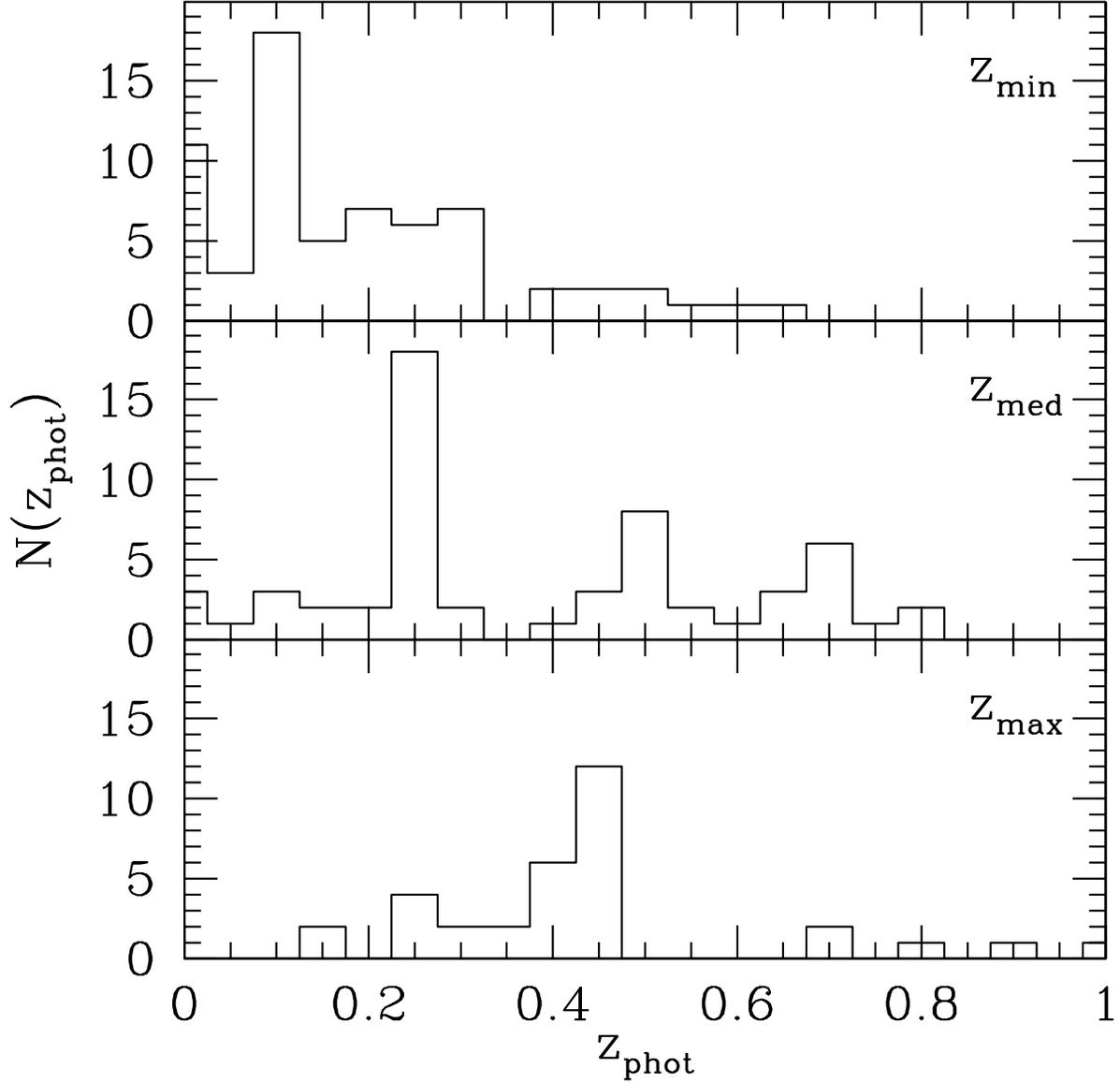}
\caption{\label{zphothist}Histograms of photometric redshifts of unresolved
sources in Abell 1689, taken from Coe et al.~\cite{Coe04}. $z_{min}$,
$z_{med}$, and $z_{max}$ refer to the minimum (2$\sigma$), medium and
maximum (2$\sigma$)
photometric
redshift obtained with the Bayesian approach described in Benitez
\cite{Benite00}).}
\end{figure}

The $z_{\rm phot}=0.5$ limit between cluster and background objects
is clear by inspecting Fig.~\ref{zphothist}. It shows a histogram of the
photometric redshift distribution of sources in the field of view of the Abell
1689 ACS image (Coe et al.~\cite{Coe04}). Apart from the most probable
photometric redshift $z_{med}$, also the smallest and largest possible
redshifts $z_{min}$ and $z_{max}$ are shown (2 sigma). Apparently there is a
pronounced peak around $z_{med}=$ 0.25, corresponding roughly to the redshift
of Abell 1689. The distributions of $z_{min}$ and $z_{max}$ show, however,
that photometric redshift values much closer to 0 and also up to about 0.5 are
possible for objects whose $z_{med}\simeq$ 0.25. We therefore adopt a limit of
$z_{med}\le0.5$ for cluster membership assignment based on photometric
redshifts, providing an upper limit on the number of cluster members. The
spectroscopic survey of Abell 1689 by Duc et al.~\cite{Duc02}, in particular
Fig.~1 in that paper, indicates that the redshift space behind the cluster is
very sparsely populated until at least z=1. The ratio of background to cluster
galaxies is only a few percent. The same holds for the number of foreground
galaxies. Therefore, the limit $z_{med}<0.5$ should yield a fair estimate of
the real number of cluster members. It is important to note here that the use
of photometric redshifts helps defining the color window only for $i\le 27$
UCD candidates, as fainter than that no photometric redshifts are available.

In $(g-i)$, the limit between $z_{\rm phot}\le0.5$ and $z_{\rm phot}>0.5$ is
well defined at about $(g-i)=1.8$ mag. This color is hence adopted as the red
limit for UCD candidates. 
The red color limit of UCDs in Fornax is $(V-I)=1.30$ mag. 
Using the ACS bandpass calibrations from Sirianni et al.~\cite{Sirian04}, this corresponds to
$(g-i)=1.77$ mag, almost identical to the adopted red color
limit based on photometric redshifts.
The blue color limit is adopted as
$(g-i)=1.2$ mag, as this is the approximate blue limit of point sources with
reliable colors. The objects bluer than that are dominated by fainter
sources with large color errors, see Fig.~\ref{cmdall}.
In total, this yields the color window $1.2<(g-i)<1.8$ mag for UCD
candidates. In $(r-z)$, analogous considerations yield a color window
$0.38<(r-z)<0.73$. The color selection
windows are quadratically broadened by the object's color error for each
object. 
The final selection criteria for UCD candidates are: SExtractor
star-classifier
larger than 0.6, $i<28$ mag,
$1.2<(g-i)<1.8$ mag, $0.38<(r-z)<0.73$ and distance from the cluster center
$r<92''$.

For objects with photometric redshift that fall into the color selection
windows, the background contamination can be estimated. There are 33 objects
with $z_{\rm phot}$ in the color window, spanning the magnitude range
$22<i<27.5$ mag. Out of these, 17 have $z_{\rm
phot}\le0.5$. This corresponds to a background contamination of 48.5 $\pm$
12\%.

\subsection{Background contamination from radial density distribution}
\label{raddens}

\begin{figure}
\plotone{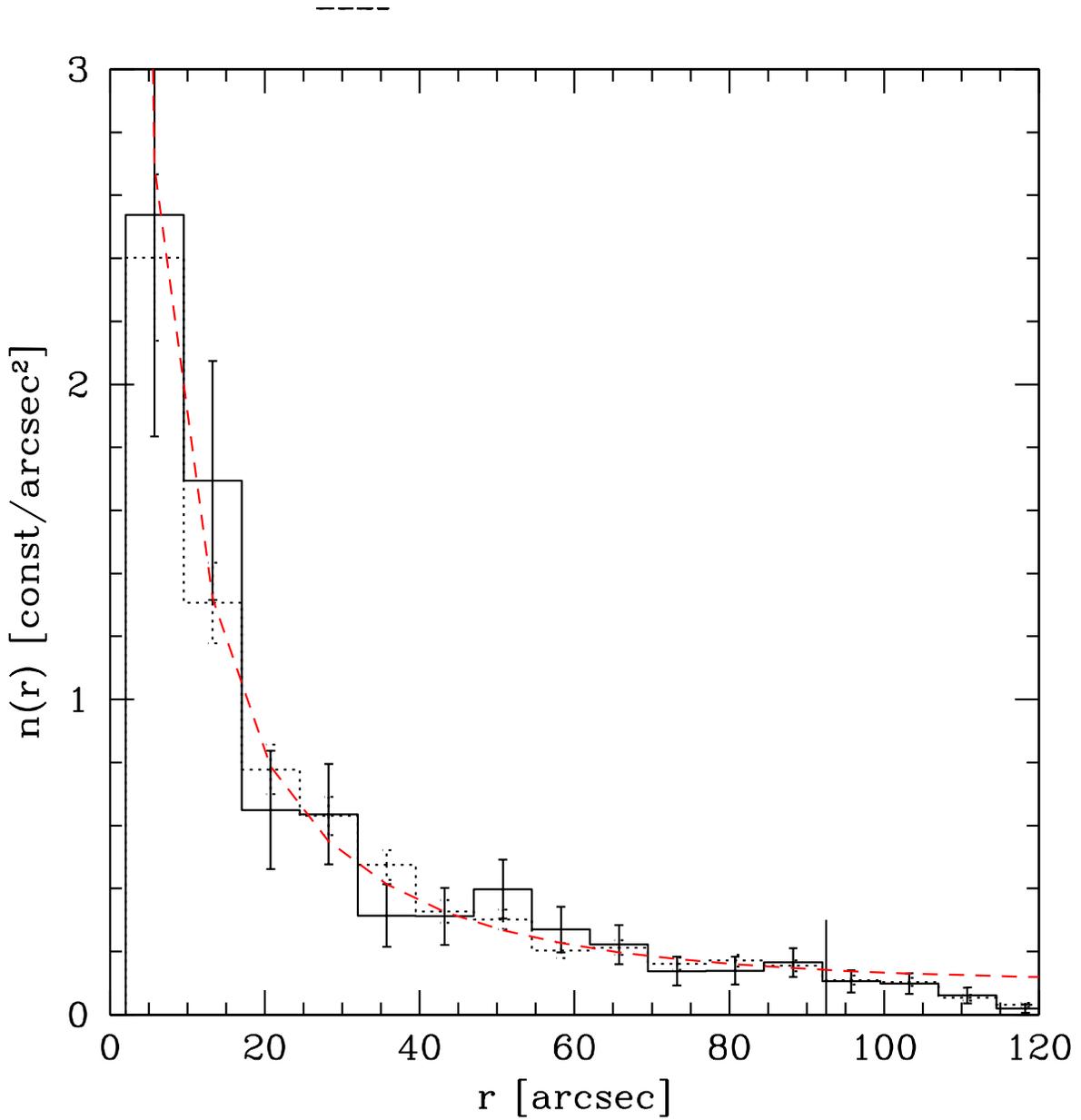}
\caption{\label{allrad}Surface density distribution of unresolved sources in
Abell 1689, arbitrary scale, plotted vs projected distance $r$ to the central
galaxy. Dotted histogram: all unresolved sources, which are Abell 1689
globular clusters in their majority (Blakeslee et al.~\cite{Blakes03}). Solid
histogram: UCD candidates. Both histograms are scaled to the total number of
objects included times an arbitrary factor. The tick at 92 $''$ marks the
image limit. The dashed 
line indicates the fit to the solid histogram, consisting of a Sersic
profile plus the background density corrected for lensing magnification
effects
(see text for further details).}
\end{figure}

\noindent Fig.~\ref{allrad} shows the surface density distribution of the UCD
candidates in Abell 1689 as selected in the previous section, together with the distribution
of all unresolved sources. In total, there are 160 UCD candidates within a
radius of 92$''$, which is the distance
from the cluster center to the image limit. 
The UCD candidates are strongly clustered, 
with their distribution agreeing very well with that
of all unresolved sources. Note that the latter objects are dominated by Abell
1689's globular cluster system (Blakeslee et al.~\cite{Blakes03}).

The total number of 160 UCD candidates in Abell 1689
must be corrected for the contribution of background number counts. To do
so, a Sersic-profile plus a radius-dependent background density was fit 
to the surface density distribution of
UCD candidates. Note that a Sersic-profile yielded a better fit than a
power-law. The inclusion of a radius-dependent background density 
takes into account that the lens magnification 
changes the number density of observed background objects. 
Following Blakeslee~\cite{Blakes99} and King et al.~\cite{King02}, we adopt
a simple isothermal profile for the magnification $\mu(\theta)$, yielding
\begin{equation}
\mu(\theta)=(1-(\frac{\theta}{\theta_E})^{-1})^{-1}
\end{equation}

We adopt $\theta_E=36 ``$ as the Einstein radius of
Abell 1689, which is the mean of the values derived by King et
al.~\cite{King02} from weak lensing and Broadhurst et al.~\cite{Broadh04}
from strong lensing. Given the undistorted background number density
$N_{bg}(m)$, it holds for the observed background number density 
$N^*_{bg}(m,\theta)$:
\begin{equation}
N^*_{bg}(m,\theta)=\frac{1}{\mu(\theta)}N_{bg}(m+2.5 \log{\mu(\theta)})
\end{equation}

Assuming a power-law distribution $N_{bg}(m)\propto 10^{m \beta}$ with
$\beta=0.32$ (Benitez et al.~\cite{Benite04}) and applying
equations
(6) and (7) of Blakeslee~\cite{Blakes99} then yields
\begin{equation}
N^*_{bg}(m,\theta)=N_{bg}(m) \times (1-(\frac{\theta}{\theta_E})^{-1})^{0.2}
\end{equation}

The fact that the exponent is positive shows that the decrease of surface
density of background sources dominates over the number count 
increase of detectable objects due to the magnification. In other words, 
the density of background objects decreases towards the cluster center.
In this case, by about 10\% as compared to a constant, radially independent background 
density.

The result of the fit is that the total background contamination for r$<$92$''$ is
26 $\pm$ 22 \%. This means that taking into account the effect of lensing on
the number density of background objects increases the number of cluster
members by about 3\%. The error of the background contamination
is considerable due to the fact
that
four variables (central surface density, scale radius, Sersic index,
background density) are fit at the same time. When fixing the fitted values
for scale radius and Sersic index, the error of the background contamination
decreases from 22\% to 13\%. In either case, the result of 26\% $\pm$ 23\% (or
$\pm 13$\%) agrees to within its errors with the background contamination
of 48.5 $\pm$ 12\% estimated from photometric
redshifts in the previous section. Adopting the mean of both values,
i.e. 37\%, about 100 out of the 160 UCD candidates should be
cluster members.
This means that there are about 7 times more
objects in the color-magnitude range of UCDs than in Fornax, where 14 UCDs
are found (as Mieske et al.
\cite{Mieske04}). A further discussion
of this is given in Sects.~\ref{radUCDdwarfs} and ~\ref{discussion}.

\subsection{Luminosity distribution of UCD candidates}
\label{lumdis}

\begin{figure}
\plotone{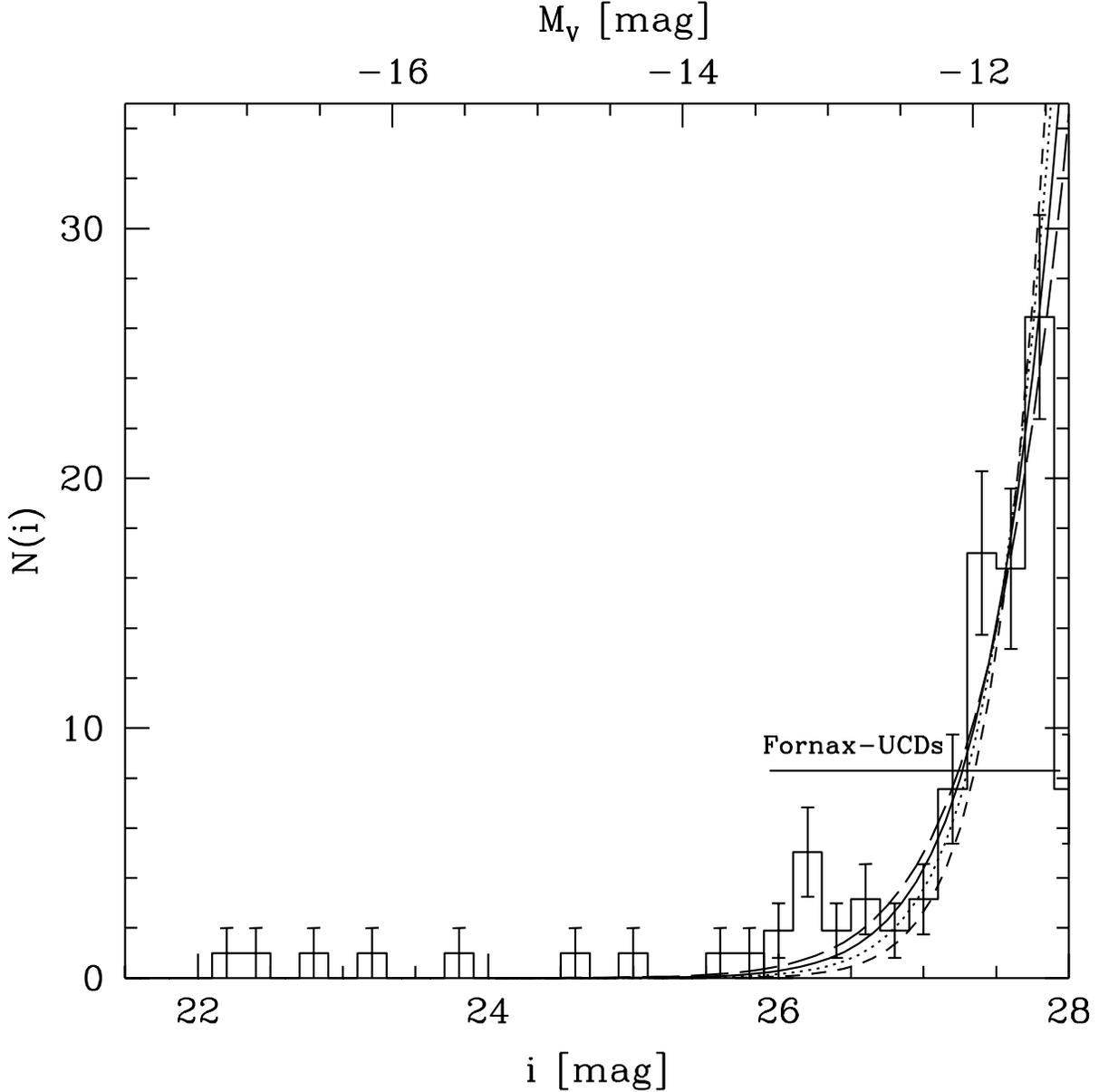}
\caption{\label{gcslf}Luminosity distribution of UCD candidates in Abell 1689
within 92$''$ of the cluster center. 
Note that for $i>26$ mag, this is the
luminosity distribution of all UCD candidates multiplied by (1-0.37), with
0.37 being the background contamination fraction found in Sect.~\ref{ncalc}.
For $i<26$ mag, all UCD candidates have a photometric redshift available.
Therefore, instead of making a statistical background decontamination, only the UCD
candidates with $z_{\rm phot}<0.5$ are included in the luminosity distribution
(see Sect.~\ref{ncalc}). The magnitude
range of the UCDs in Fornax is indicated by the horizontal tick.
The long dashed, solid, dotted, short dashed lines
correspond to a Gaussian GCLF at Abell 1689's distance with $\sigma=$ 1.5,
1.4, 1.3, 1.2 mag respectively.  The fitted respective total number of GCs is
$N_{\rm glob}=$ 17100, 30100, 59300, 135400. The fit with $\sigma=$ 1.4 mag and
corresponding $N_{\rm glob}=$ 30100 has the lowest $\chi ^2$. }
\end{figure}

\noindent Fig.~\ref{gcslf} shows the luminosity distribution of UCD candidates
in Abell 1689. The counts have been corrected for the contribution of background sources
using the value of 37\% contamination derived in the previous
Sect.~\ref{raddens}. Gaussian GCLFs with
width $\sigma=1.2$, 1.3, 1.4 and 1.5 mag all match reasonably well the number
counts for $i \le$ 26.8 mag ($M_V \simeq$ -12.7 mag), resulting in a total
number $N_{\rm glob}$ between 135400 for $\sigma=1.2$ and 17100 for $\sigma=1.5$ mag.
The $\chi ^2$ value of the fit is best for $\sigma=1.4$ mag, consistent with
what is usually found for GC systems (Kundu \& Whitmore~\cite{Kundu01}). Note
that unlike for the Fornax case, where the GC system of NGC 1399 has been
investigated beyond the turnover magnitude (TOM), this is not possible for
Abell 1689. In $i$, the TOM lies at about $i\simeq$ 31.8 mag, far beyond the
reach of even the ACS. This is why the total number of GCs can be restricted
only poorly with the existing data. Crowding is not a significant effect:
random position simulations of globular clusters belonging to a GCS
 with TOM $i=31.8$ mag, $\sigma=1.4$ mag and $N_{\rm glob}=50000$ show that only 1.5\% of the clusters
with $i<29.5$ mag -- ~ 1 mag fainter than the completeness limit -- are
less than 2 FWHM away from their next neighbor.

For magnitudes brighter than $i\simeq$ 26.8 mag ($M_V\simeq -12.7$ mag), there
is an overpopulation of UCD candidates with respect to any of the adopted
Gaussian LFs. The number of GCs with $\le 26.8$ mag expected from the GCLF
with width $\sigma=1.4$ mag is 4.8, while the number of candidate UCDs is
20.9. The two values are inconsistent at the 3.5 $\sigma$ level. Note that the
non-integer numbers are caused by the statistical background decontamination.

The faint magnitude limit of $M_V\simeq -12.7$ mag for the overpopulation is
about 1 mag brighter than the faint magnitude limit of UCDs in Fornax. This might
indicate that the objects equivalent to UCDs in Fornax are 
brighter in Abell 1689. Before making that conclusion, it must be clarified
whether the several $10^4$ to $10^5$ GCs that are needed to explain the number
counts for $i>26.8$ mag, can at all be contained in the GC system of Abell
1689. Assuming $S_N=$ 6, a value found for example for the Fornax cluster's
central galaxy NGC 1399 (Dirsch et al.~\cite{Dirsch03}), the total number of
GCs belonging to the ten brightest cluster galaxies in Abell 1689
($-23.1<M_V<-21.4$ mag, Zekser et al.~\cite{Zekser04}) is 44500, being of the
order of the values found above. Thus, it is not necessary to introduce other
objects than genuine GCs contributing to the number counts for $i\ge$ 26.8
mag.

Therefore, from the magnitude distribution we conclude that for $i\le$ 26.8
mag we find evidence for an overpopulation with respect to the GCLF, while for
$i\ge$ 26.8 mag the number counts are consistent with a regularly rich GCS
following a Gaussian LF.

\subsection{Color distribution of UCD candidates vs. normal dwarf galaxies}

\begin{figure}
\plotone{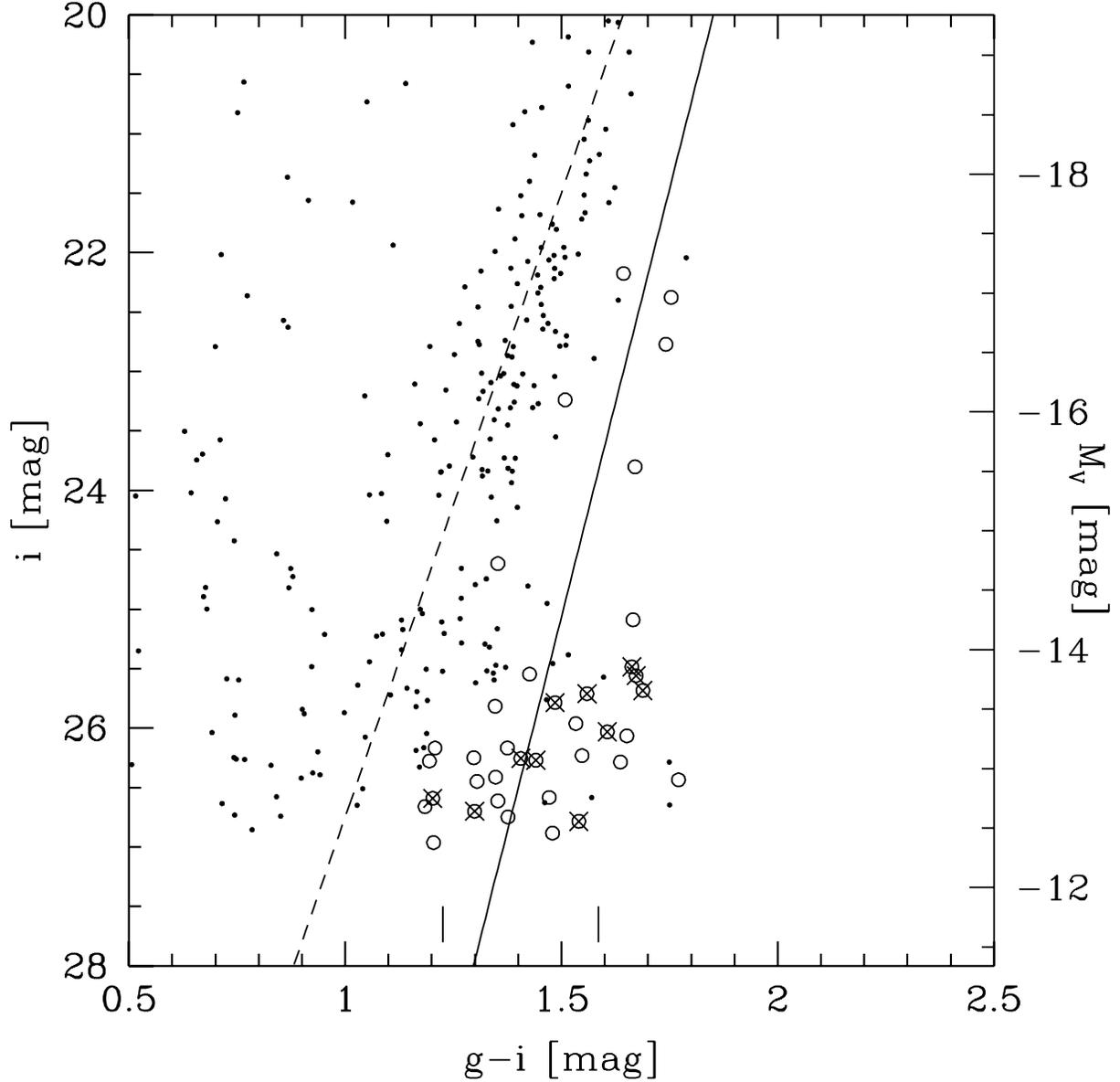}
\caption{\label{gcscmdgibright}CMD in $i$, $(g-i)$ of UCD candidates with
$i<27$ mag, indicated as circles, and resolved objects in Abell 1689 with
$z_{\rm phot}<0.5$ taken from Infante et al.~\cite{Infant04}, indicated as
dots. Crosses mark unresolved UCD candidates with $z_{\rm phot}>0.5$. The
lines indicate the fitted CM relation to the resolved sources (dashed) and the
UCD candidates (solid). Note that the brightest UCD in Fornax has $M_V=-13.4$
mag, several mag fainter than found here. The vertical ticks denote the
average position of the blue and red peak of the GCLF (Kundu \& Whitmore
\cite{Kundu01}).}
\end{figure}

\begin{figure}
\plotone{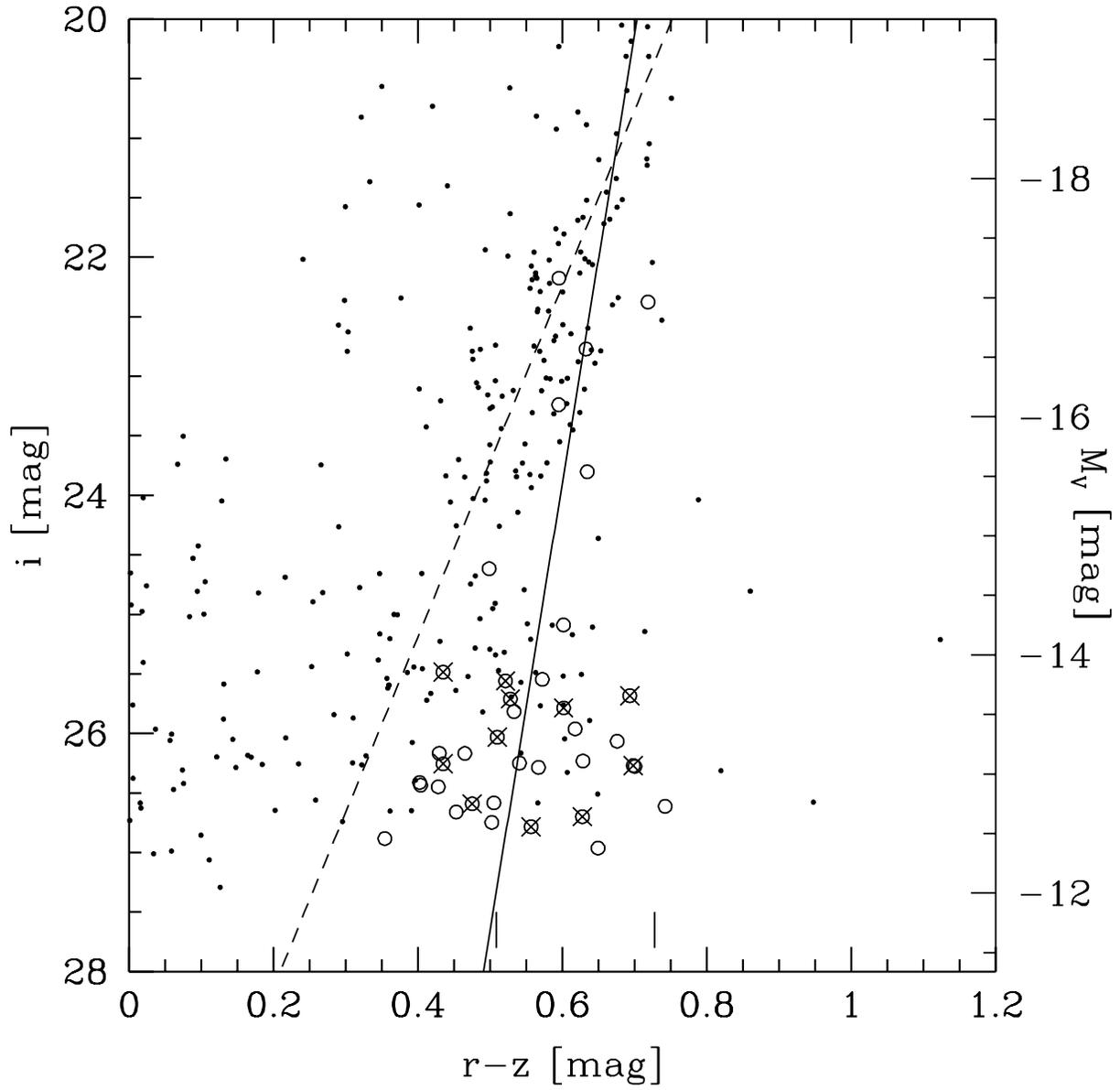}
\caption{\label{gcscmdrzbright}CMD in $i$, $(r-z)$ of the same objects as in Fig.~\ref{gcscmdgibright}.
Legend analogous to Fig.~\ref{gcscmdgibright}.}
\end{figure}

\noindent In order to investigate the sub-population of UCD candidates which
constitute an overpopulation with respect to Abell 1689's GCLF,
Figs.~\ref{gcscmdgibright} and~\ref{gcscmdrzbright} show CMDs of UCD
candidates with $i<27$ mag in $(g-i)$ and $(r-z)$, respectively. The colors of
resolved objects with $z_{\rm phot}<0.5$ are given to illustrate the location
of the CM sequence for dwarf galaxies in the cluster. A detailed paper on these
resolved objects is in preparation (Infante et al.~\cite{Infant04}).

The following main features are extracted from the CMDs:

\begin{enumerate}

\item UCD candidates extend to much brighter magnitudes than in Fornax. The
brightest UCD candidate in Abell 1689 has $M_V\simeq$ -17 mag, about as bright
as the cE M32 (Mateo \cite{Mateo98}).

\item UCD candidates and dwarf galaxies follow color-magnitude trends in the sense that
color becomes redder with growing luminosity. For the UCD candidates, the
slope is -0.069 $\pm$ 0.006 for $(g-i)$ and -0.027 $\pm$ 0.005 for $(r-z)$.

\item UCD candidates are redder than dwarf galaxies of the same brightness, a
behavior which is more pronounced in $(g-i)$ (about 0.3 mag difference) than
in $(r-z)$ (between 0.0 and 0.2 mag difference).

\item UCD candidates are more consistent with the red than the blue peak of
the bimodal GC color distribution.

\end{enumerate}

Features 2, 3 and 4 are qualitatively consistent with what has been observed
for UCDs in the Fornax cluster (Mieske et al~\cite{Mieske04}). The
interpretation is that UCDs are created by tidally disrupting nucleated
dwarf galaxies (dE,Ns), which
after some passages through the cluster's potential well can only retain the
nuclear part, hence become ``ultra compact'' (Bekki et al.~\cite{Bekki03}). In
a CMD a dE,N moves towards fainter magnitudes at unchanged color. The final
products of this threshing procedure, namely the UCDs, then define a CM
relation with similar slope to that of dEs, but shifted red-wards. The most
interesting feature is that the Abell 1689 UCD candidates reach the
luminosity of very bright ``normal'' dwarf galaxies: they extend
to $M_V\simeq-$ 17 mag, while the magnitude limit between  dwarf and giant galaxies
is only 1-2 mag brighter (Hilker et al.~\cite{Hilker03}).
This could mean that at least in the
case of the brightest UCD candidates we are seeing objects in the transition
phase between dE,Ns and UCDs, as a magnitude difference of about 4 mag between
 dE,N and UCD is expected (Bekki et al.~\cite{Bekki03}).

\begin{figure}
\plotone{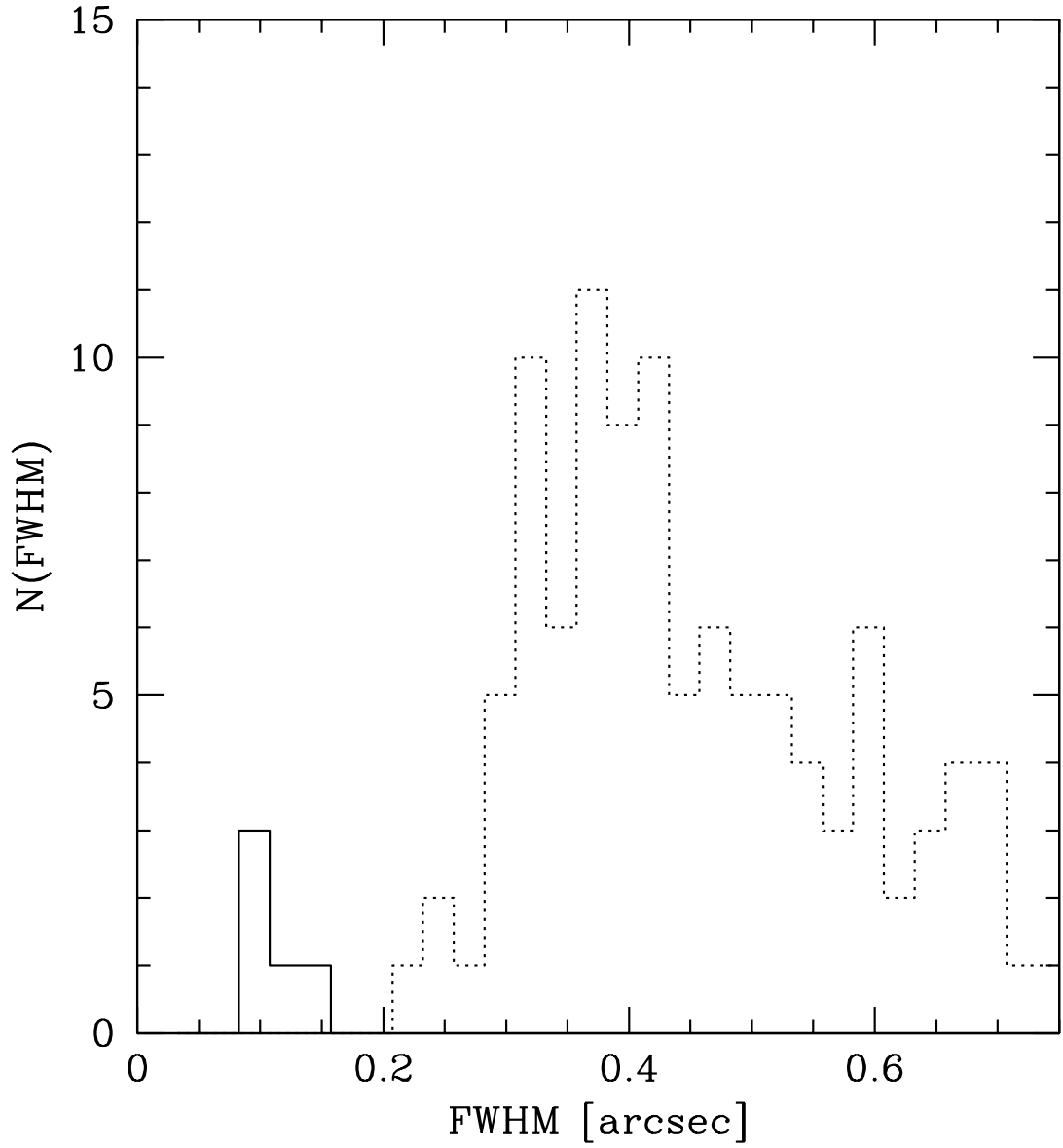}
\caption{\label{gcsdEsizehist}Size distribution of objects from Fig.~\ref{gcscmdgibright} with $22<i<24$ mag.
Solid histogram: UCD candidates. Dotted histogram: resolved objects, taken
from Infante et al.~\cite{Infant04}.}
\end{figure}

The idea that the brightest UCD candidates are just the smallest genuine dEs
of a continuous distribution is another possible explanation for our findings.
However, Fig.~\ref{gcsdEsizehist} shows that the size distribution of all
(unresolved and resolved) objects in the magnitude range $22<i<24$ mag is not
consistent with that assumption. The size distribution of resolved sources has
a sharp lower cutoff at about 0.2$''$ FWHM, twice as large as the FWHM of the
UCD candidates. This dichotomy is as well present in the $gi$ CMD in
Fig.~\ref{gcscmdgibright}, where the UCDs are well separated from the resolved
objects.

In Sect.~\ref{discussion}, the implications of our findings will be further
discussed.

\subsection{Radial distribution of UCD candidates vs. normal dwarf galaxies}
\label{radUCDdwarfs}
\begin{figure}
\plotone{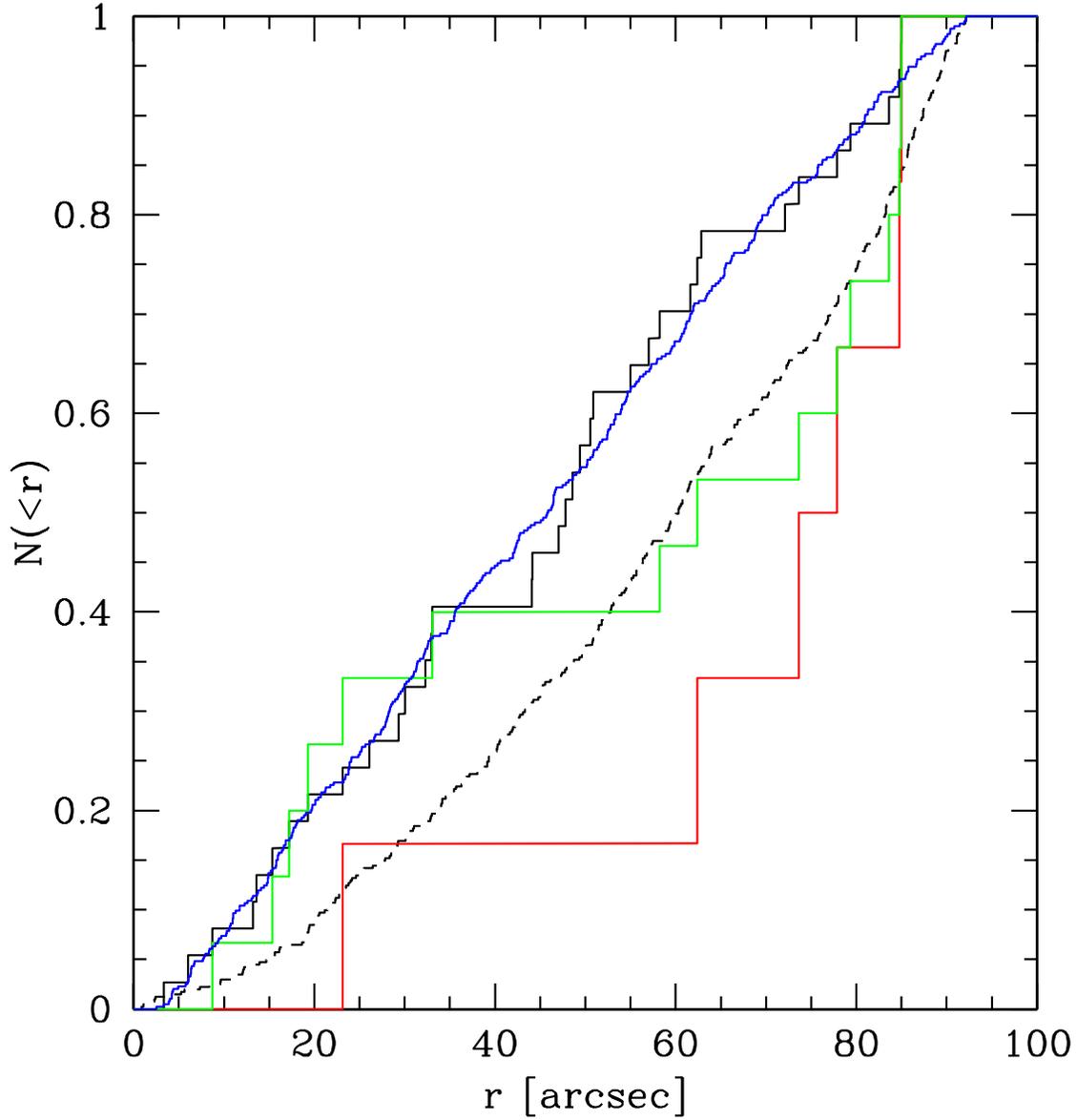}
\caption{\label{radcumgcsdE}Cumulative radial distribution  of objects in
Abell 1689, restricted to the field of view of 92$''$ radius.
Blue histogram: UCD candidates with $i>27$ mag. 
Black histogram: UCD candidates with $i<27$ mag. Green histogram: UCD
candidates with $i<26$ mag. Red histogram: UCD candidates with $i<25$ mag.
Dashed histogram: dwarf galaxy candidates in Abell 1689 with $i<26$ mag and
$(g-i)>1.1$, taken from Infante et al.~\cite{Infant04}.}
\end{figure}

\noindent Fig.~\ref{radcumgcsdE} shows the cumulative radial distribution of
UCD candidates and dE candidates in Abell 1689, with the latter data taken
from Infante et al.~\cite{Infant04}. The sample of UCD candidates is split
into three overlapping sub-samples: $i<27$ mag (UCD$_1$ hereafter), $i<26$ mag (UCD$_2$)
and $i<25$ mag (UCD$_3$). In addition, the distribution of
unresolved objects with $i>27$ mag in the color range of
UCD candidates (GCS hereafter) is indicated. Note that
the sample GCS contains the UCD candidates with $27<i<28$ mag and also all
fainter sources and is therefore dominated by genuine globular clusters,
according to the results of Sect.~\ref{lumdis}.

From Fig.~\ref{radcumgcsdE} it is evident that the UCDs
with $i<27$ mag are more strongly clustered than the two brighter sub-samples
and also the dEs. A KS test shows that the cumulative radial distribution of
UCD$_1$ is drawn from the same distribution than GCS at 97\% confidence. The
distribution of UCD$_1$ is inconsistent with that of the dEs at the 98\%
confidence level. This disagreement with the dEs drops to 53\% for UCD$_2$ and
68\% for UCD$_3$. UCD$_1$ is inconsistent with UCD$_2$ at 77\% and with
UCD$_3$ at the 95\% confidence level. For the six objects contained in
UCD$_3$, no clustering is detectable.

These findings show that {\it only} for $i<26$ ($M_V<-13.4$) mag, UCD
candidates show a different radial distribution compared to the globular
cluster system of Abell 1689. The fact that their distribution agrees better
with that of the dwarf galaxies is consistent with the threshing scenario.
Going back to the luminosity distribution of UCD candidates as shown in
Fig.~\ref{gcslf}, it is found that in the magnitude range $26<i<26.8$ mag, the
predicted number of GCs from the best fit GCLF is 4.3, while the number of UCD
candidates is 12.1, inconsistent at the 2.2 $\sigma$ level. For $i<26$ mag, the
disagreement is much stronger: only between 0 and 1 GC are expected from the
GCLF, compared to 10 actually found objects.

\noindent In conclusion, UCD candidates are well separated in luminosity and spatial
distribution from GCs for $i<26$ ($M_V<-13.4$) mag. For $i>26$ ($M_V>-13.4$)
mag, our data suggest that genuine GCs are an important fraction. UCD
candidates in that magnitude regime cannot be distinguished from GCs, and they
might mix up with them. A minimum number of about ten UCDs with $M_V<-13.4$ is
then consistent with our results.

\section{Discussion}
\label{discussion}

\noindent The aim of this section is twofold: first, compare the expected
number of stripped dE,N nuclei in Abell 1689 with the number of UCD
candidates derived in the previous section. Second, discuss several 
alternative possibilities for the
identity of the very bright UCD candidates.
\subsection{Number of UCD candidates}

\noindent Based on their simulations, Bekki et al.~\cite{Bekki03} predict the
number of UCDs expected from threshing dE,Ns for the case of the Fornax and
the Virgo cluster. For Abell 1689 no such simulations exist, yet.
Abell 1689 has a several times higher x-ray temperature than Virgo 
(Young et al.~\cite{Young02}, Xue \& Wu~\cite{Xue02}) and about half as many
spirals per E/S0 galaxy (Ferguson~\cite{Fergus89}, Balogh et al.~\cite{Balogh02}).
However, the two clusters are similar in terms of enclosed
mass and size, two factors which predominantly determine the number of UCDs
expected from threshing dE,Ns. While $M=5\times 10^{14} M_*$, $M/L=500$ and a
scale radius $r_s=226$ kpc for Virgo (Bekki et al.~\cite{Bekki03}),
$M\simeq 1 \times 10^{15} M_*$, $M/L\simeq 400$ and $r_s\simeq 350$ kpc for Abell 1689
(Broadhurst et al.~\cite{Broadh04}, King et
al.~\cite{King02}, Tyson \& Fischer~\cite{Tyson95}). The mass value
for Abell 1689 differs somewhat depending on the method used to
derive it. The strong lensing study of Broadhurst et al. obtains 
$\simeq 2 \times 10^{15} M_*$, King et al. get $\simeq 5 \times 10^{14} M_*$ from weak
gravitational lensing, 
while current X-ray estimates
(Xue \& Wu~\cite{Xue02}, Andersson \& Madejski~\cite{Anders04})
yield lower results than the King et al. values by a factor of about 2.

For the Virgo cluster, 46
UCDs are expected with a maximum projected radius of 700 kpc (Bekki et
al.~\cite{Bekki03}). 
The number of
UCDs and their radial extension should therefore be expected similar for Abell
1689. Note that for the Fornax cluster, which has an almost ten times smaller
mass and three times smaller scale radius, Bekki et al. predict 14 UCDs, in
very good agreement with the number of UCD candidates found in Fornax by
Mieske et al.~\cite{Mieske04}.

The maximum projected radius of the present investigation is about 285 kpc.
According to Bekki et al.'s simulation for the Virgo cluster, about 50\% of
UCDs created by threshing dE,Ns are expected within this radius. Therefore, of
the order of 20-25 UCDs should be found in the ACS image of Abell 1689. As
shown in Sect.~\ref{lumdis}, only for $i<26$ mag ($M_V<-13.4$ mag) is it
possible to reasonably estimate the number of UCD candidates, as they are
distributed spatially more extended and are brighter than expected for even
the brightest GCs of Abell 1689's GCS. In that magnitude range, 10 UCD
candidates are found. This number constitutes, of course, a lower limit on the
number of UCDs. The overpopulation found for $26<i<26.8$ mag with respect to
the GCLF would contribute another 6-7 objects, but note that in this magnitude
range the radial distribution of UCD candidates is indistinguishable from that
of the GCs. Fainter than $i\simeq 26.8$ mag, the majority of objects are GCs,
but there can be UCDs mixing up with them.

In total, a number of 20-25 of UCDs is consistent with our findings, which
provide a lower limit of about ten UCDs. Spectroscopic data of the (brightest)
UCD candidates is necessary in order to definitely determine their cluster
membership.

\subsection{Dwarf galaxies caught in the act of threshing?}
\label{resolved}

\noindent If all UCD candidates with $i<26$ mag really were members of Abell
1689, this would imply that there are very bright UCDs reaching luminosities
of $M_V\simeq -17.5$ mag ($i \simeq 22$ mag), almost 4 mag brighter than in
Fornax. We might therefore see dwarf galaxies that are still in the process of
disruption, being in the early stages of the threshing process as simulated by
Bekki et al.\cite{Bekki03}. The assumption that the UCD candidates originate
from dwarf galaxies rather than globular clusters is supported by the fact
that the very bright UCD candidates have a radial distribution which is more
consistent with that of Abell 1689 dwarf galaxies than with the
GCS. Moreover,
 the brightest UCD candidates are by far too luminous for globular clusters.

Another supportive finding is that the UCD candidates follow color-magnitude
trends both in $(g-i)$ and $(r-z)$, see Figs.~\ref{gcscmdgibright} and
\ref{gcscmdrzbright}, which place them redward of the dwarf galaxies. For the
very bright UCD candidates with $i<25.2$ mag, these relations are best
defined. For $(g-i)$, the slope in that magnitude regime is -0.086 $\pm$
0.012. The slope of the dwarf galaxy relation is -0.095, consistent with that
of the UCD candidates. The mean color difference between the UCD and dwarf CM
relation is 0.28 mag. Assuming that the UCD candidates are (partially)
threshed dE,Ns of unchanged color (see Mieske et al.\cite{Mieske04}), that
means that the UCD candidates are about $\frac{0.28}{(0.086+0.095)/2} \simeq$
3 mag fainter than their progenitors. In $(r-z)$, the UCD slope is -0.0315
$\pm$ 0.01 and the dwarf galaxy slope -0.068 $\pm$ 0.02, with the mean color
difference between both relations being 0.092 mag. This implies a magnitude
difference of $\frac{0.092}{(0.0315+0.068)/2} \simeq$ 2 mag between UCD
candidates and progenitor dE,Ns.

These differences are smaller than the 4.1 mag difference between UCDs and
progenitor dE,Ns predicted by Bekki et al.\cite{Bekki03} and the difference of
about 5 mag between dE,Ns and their nuclei found observationally by Lotz et
al.\cite{Lotz01}. It therefore supports the hypothesis that the very bright
UCD candidates are dE,Ns that have not yet transformed entirely to naked
nuclei, as also suggested from their higher luminosities as compared to the
Fornax cluster.

\begin{figure}
\psfig{figure=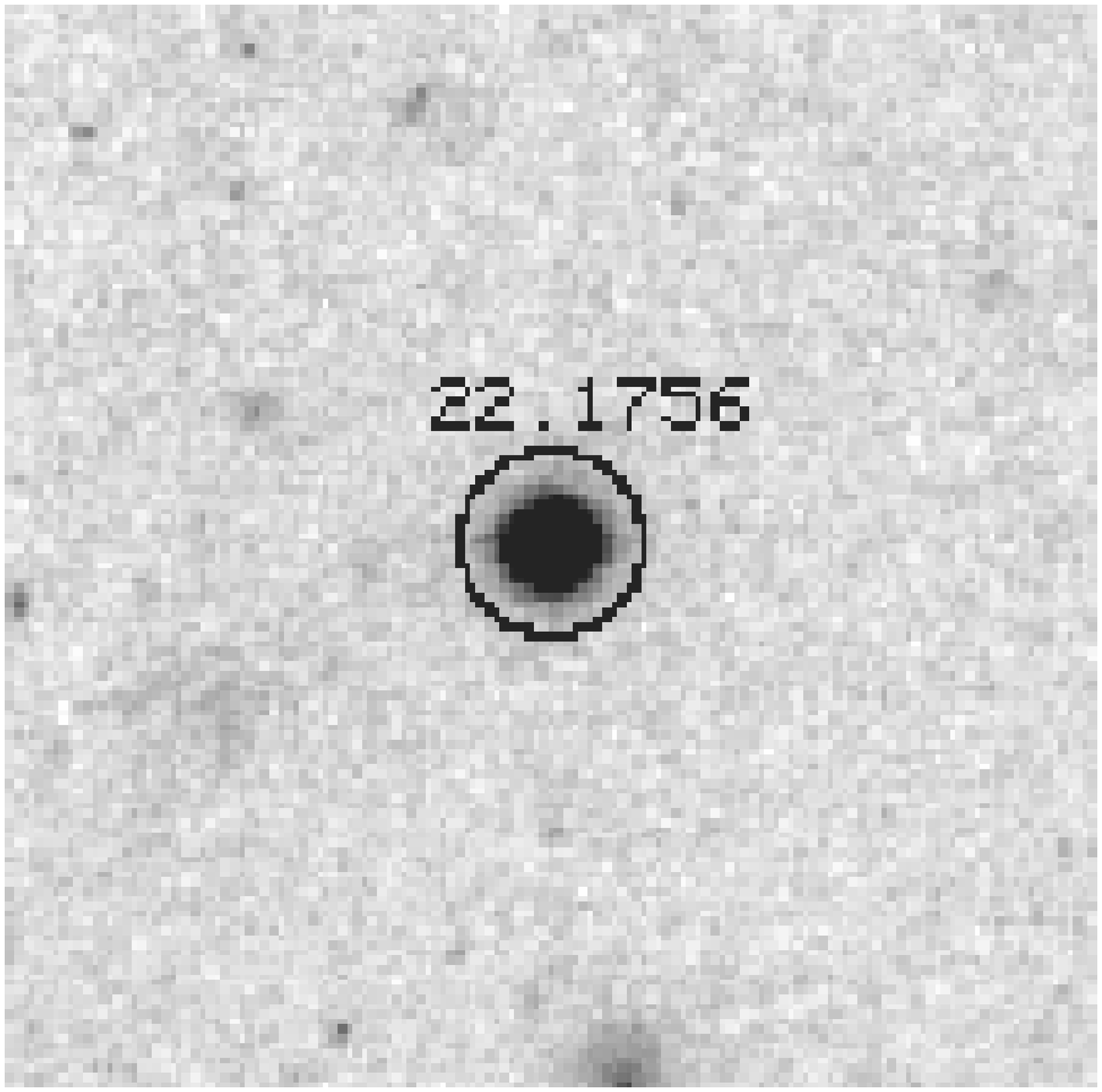,width=3.85cm}
\psfig{figure=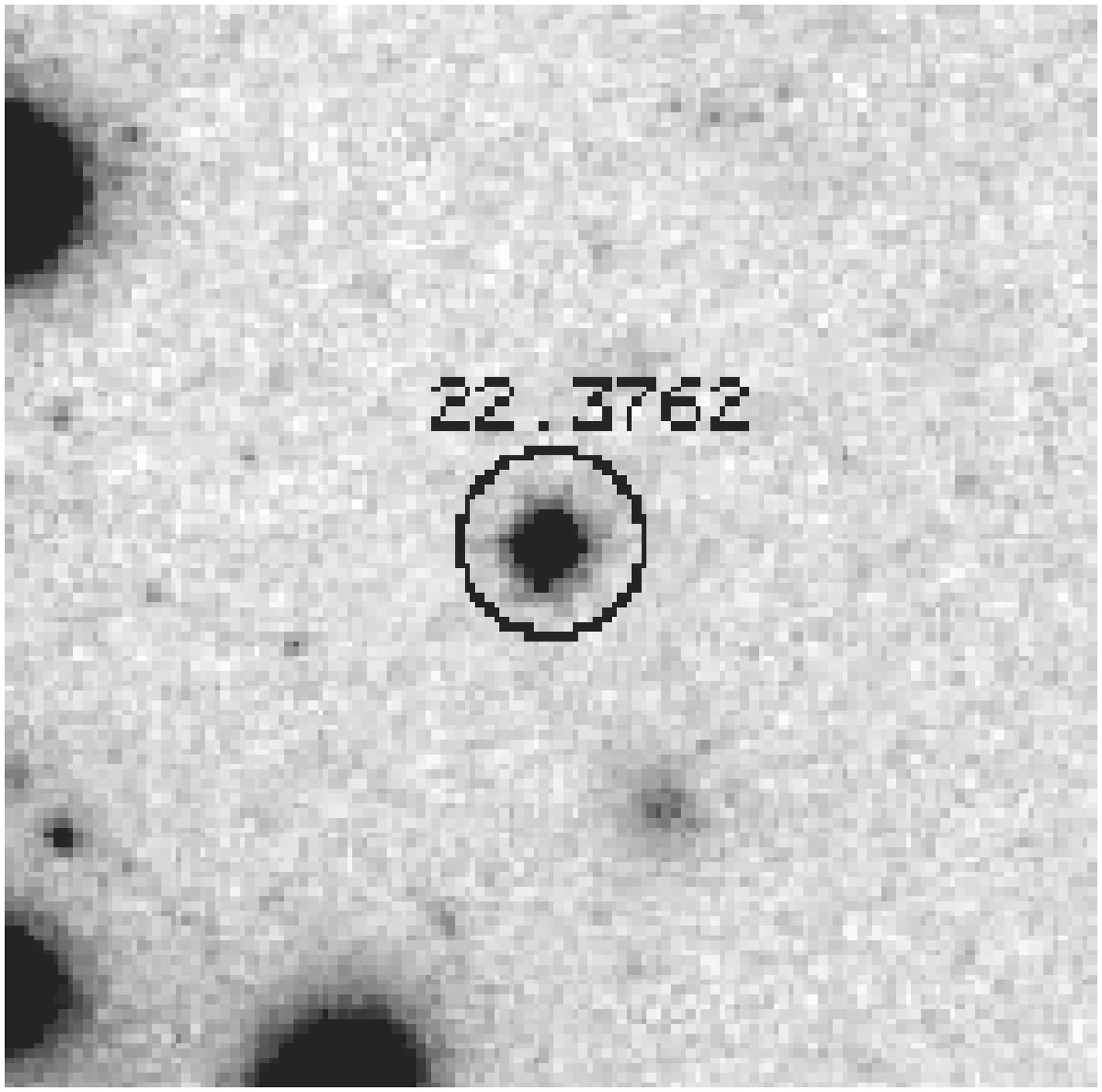,width=3.85cm}
\psfig{figure=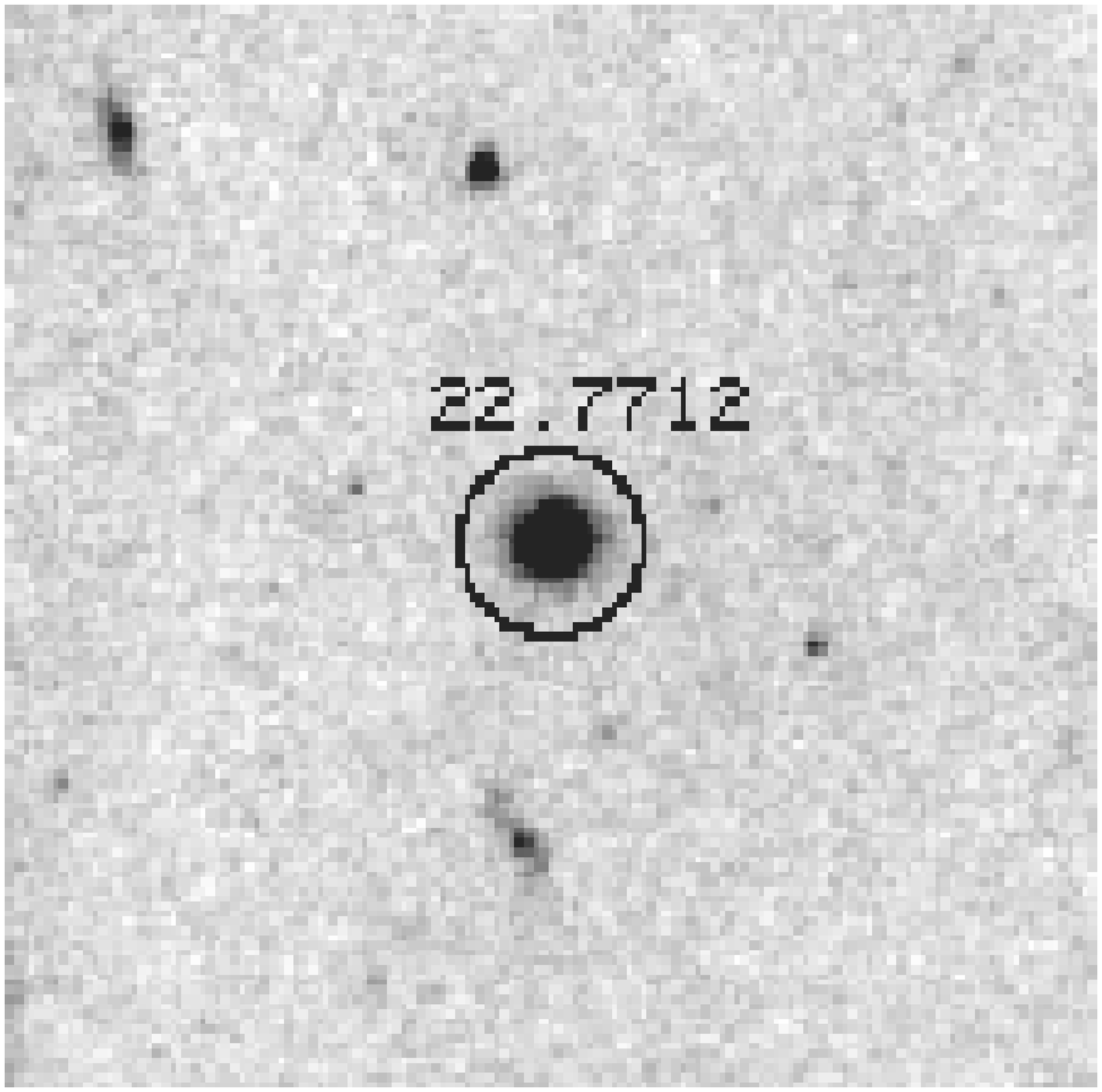,width=3.85cm}
\psfig{figure=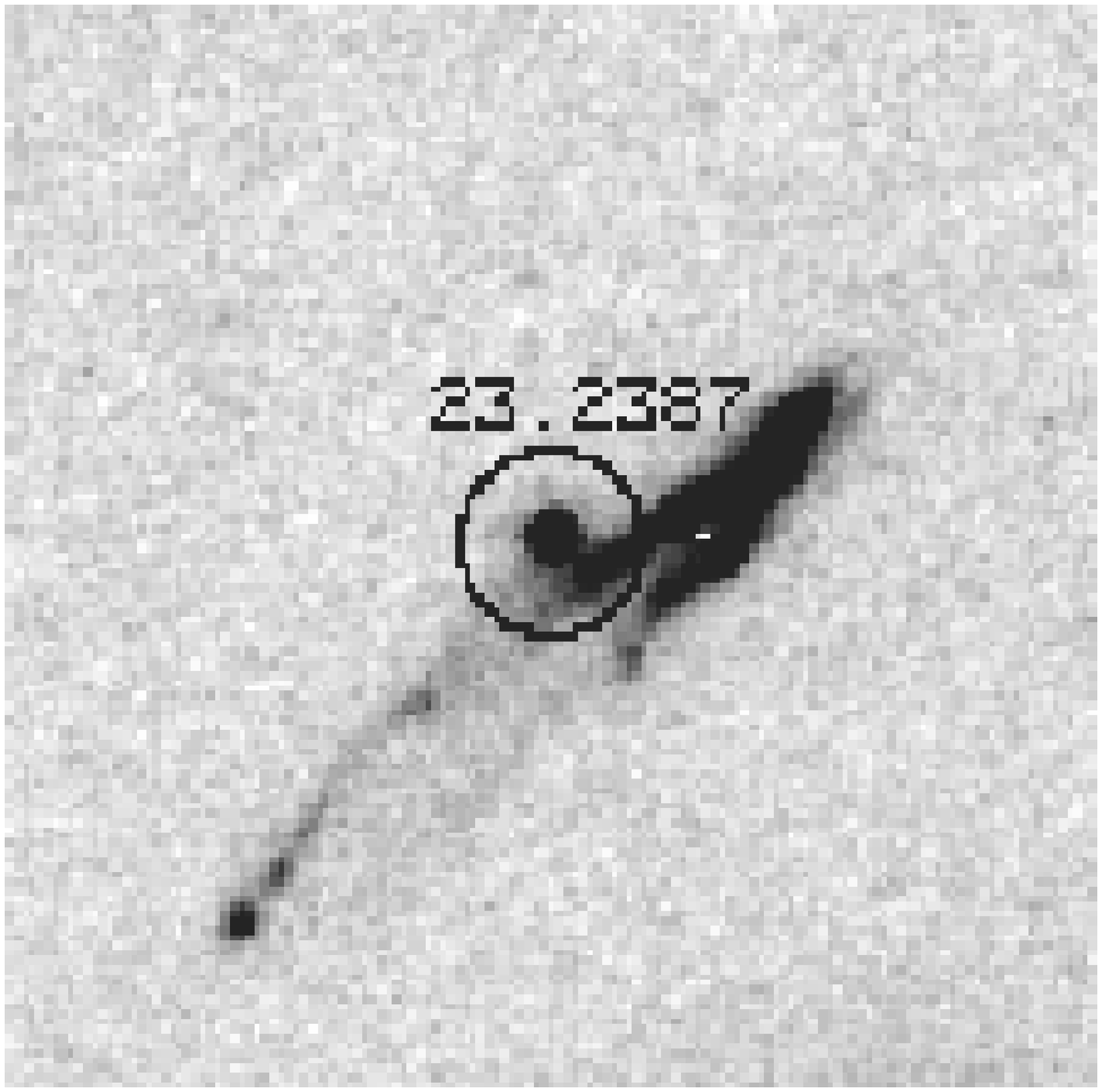,width=3.85cm}\\
\psfig{figure=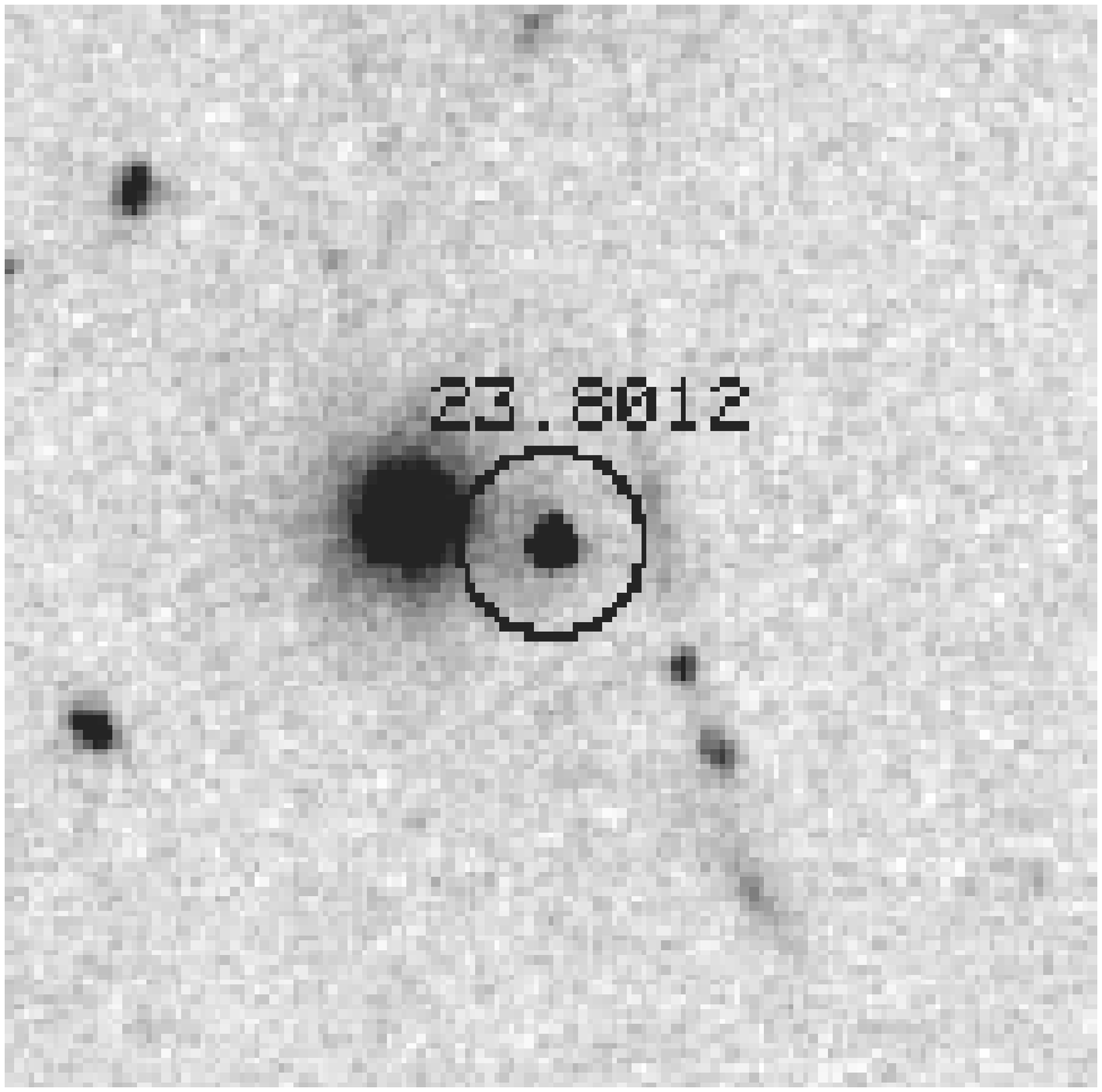,width=3.85cm}
\psfig{figure=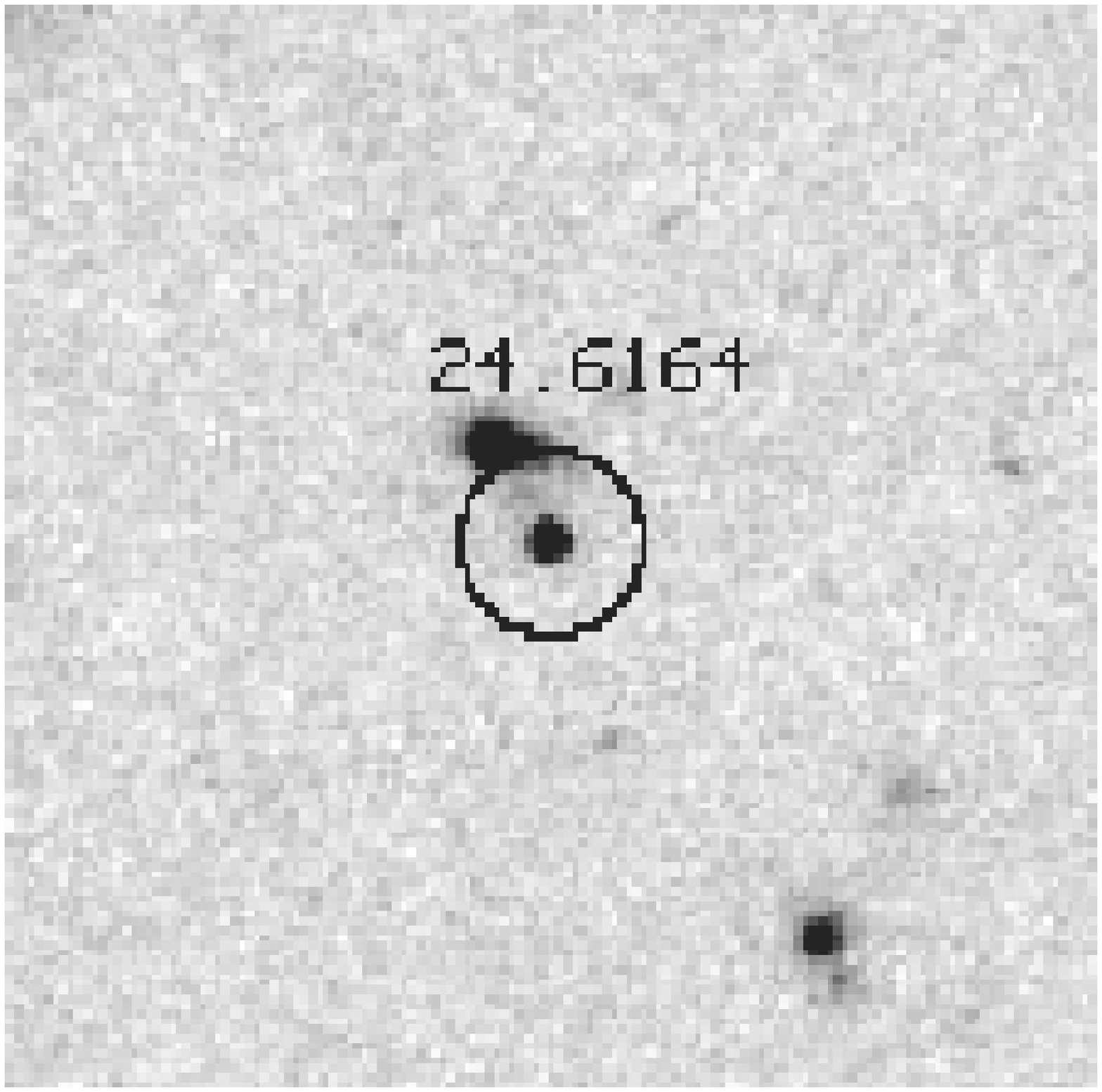,width=3.85cm}\hspace{2cm}
\psfig{figure=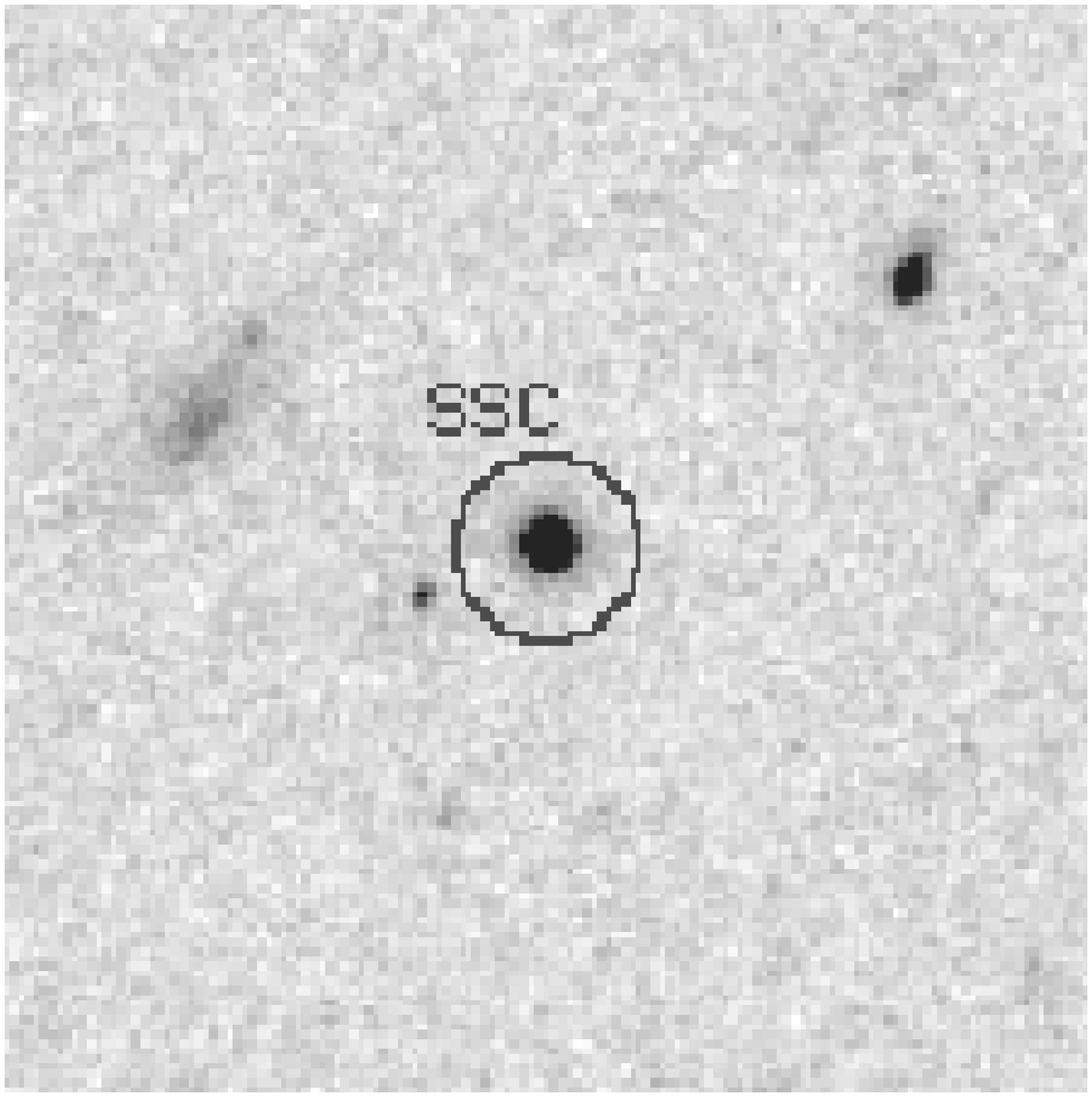,width=3.85cm}
\caption{\label{images}7$''$$\times$7$''$ thumbnails of the six UCD candidates
with $i<25$ mag plus the one stellar super-cluster (SSC) candidate with
$i\simeq 23.5$ mag, see text for details. $i$ is indicated for the UCD
candidates. The two UCD candidates with 22.17 and 22.77 mag are slightly
resolved, see Fig.~\ref{ucdbrightcog} and text.}
\end{figure}

\begin{figure}
\plottwo{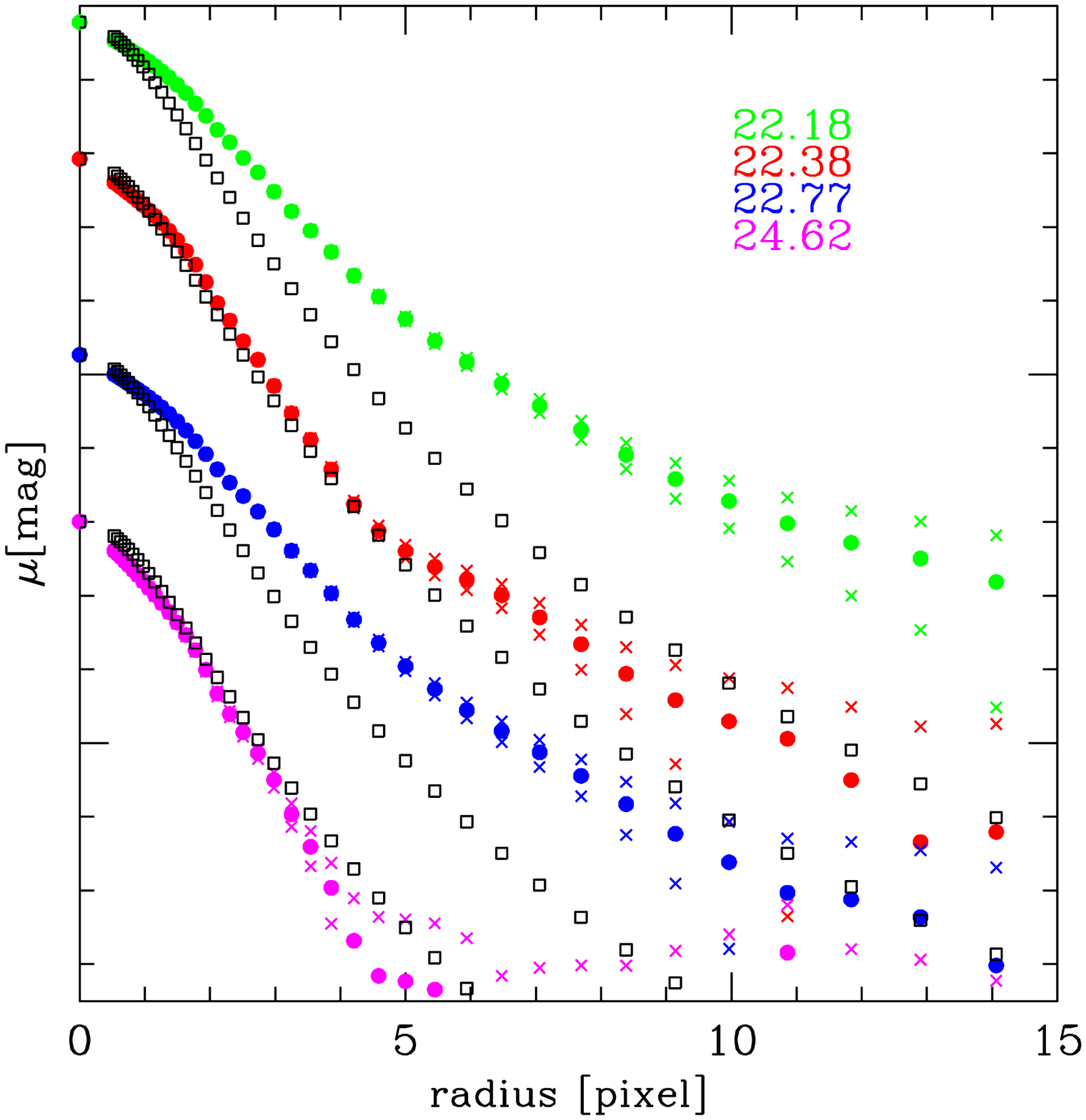}{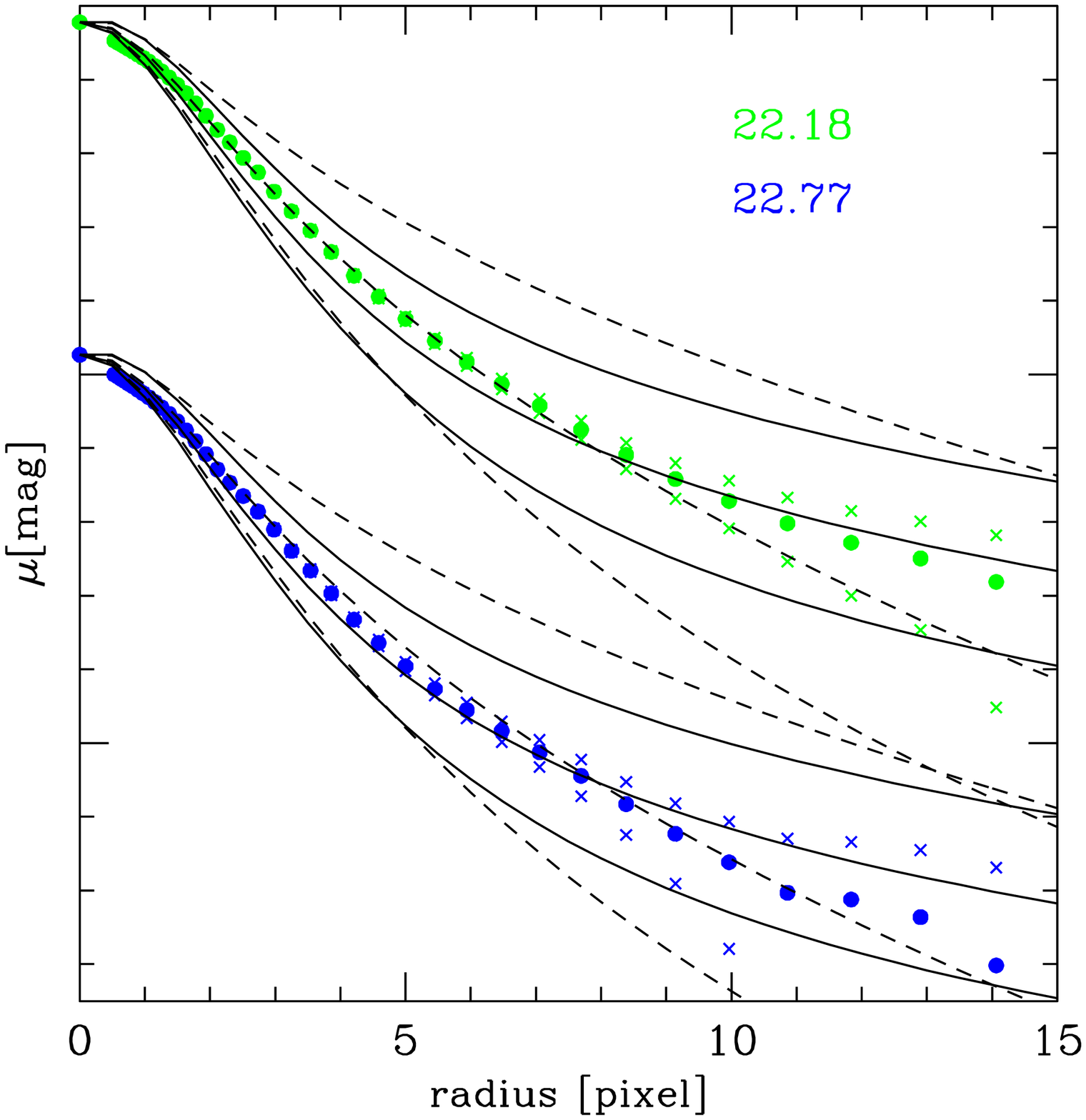}
\caption{\label{ucdbrightcog}{\bf Left panel:}
Surface brightness profile of four of the six
very bright UCD candidates shown in Fig.~\ref{images}, as measured on the
detection-image (the two lacking UCD
candidates have too close neighbors). The profiles have been scaled arbitrarily in order not to
overlap and hence allow clearer distinction; one tick on the y-axis is one mag
in surface brightness. The corresponding total magnitudes are
indicated, allowing cross-identification with Fig.~\ref{images}. Filled
circles indicate the measured profile, crosses indicate the result when adding
and subtracting one sky standard deviation. Open squares indicate a Moffat
profile with 0.11$''$ (2.2 pixel) FWHM 
(corresponding to the resolution of the image) with
the same peak intensity as the respective object. Apparently, the two sources with 22.18
and 22.77 mag are slightly resolved. {\bf Right panel:}
Surface brightness profile of the two resolved sources. The solid lines
indicate King profiles with core radii of  0.45, 0.225 and 0.1125 pixel
(top to bottom) convolved with the PSF and scaled to match the central
intensity of the observed profile. The dashed lines are convolved Sersic profiles
with n=2 and effective radii of 4, 2 and 1 pixel
(top to bottom).}
\end{figure}

It is interesting to note that $M_V \simeq -17$ mag corresponds approximately to
the total luminosity of M32, with M32's high surface brightness bulge being about 1 mag fainter
(Mateo~\cite{Mateo98}, Graham~\cite{Graham02}). 
The bulge component of M32
has an effective radius $r_{\rm eff}$ of about 100 pc (Faber et
al.~\cite{Faber89}, 
Graham~\cite{Graham02})
which at the distance of Abell 1689 corresponds to about 0.064$''$ effective
diameter.
This angular size is about half of the PSF FWHM. An M32 A1689 equivalent,
like the very bright UCD candidates, might therefore be marginally resolved.
Fig.~\ref{images} shows the six brightest UCD candidates ($i<25$ mag) in the
ACS image, with the same intensity cuts for each thumbnail.
 The two sources with 22.18 and 22.77 mag appear slightly more
extended than the object with 22.38 mag.
The SExtractor star classifier is 0.88 and 0.92 for the two more
extended sources, and between 0.93 and 0.98 for the other four ones. The
differences become more evident in Fig.~\ref{ucdbrightcog}. Here, the surface
brightness profile of the three brightest and the faintest of the six objects
is compared with the PSF profile of the ACS images. Evidently, the two
sources with 22.18 and 22.77 mag have an extended envelope
compared to the PSF profile. 
Comparing their brightness profile with
convolved
King-profiles shows that a core radius of about 0.225 pixel -- i.e. a core
diameter
and FWHM of 0.45 pixel or 0.0225$''$ -- fits the observed profile best for both
sources. This corresponds to a King-profile core radius of about 35 pc at the distance of
Abell 1689, several times larger than found for the UCDs in Fornax
(Drinkwater et al.~\cite{Drinkw03}). The King-profile can per definitionem
not be characterized by an effective radius, as its integrated intensity
rises proportional to $log(r)$ for large values of $r$. Therefore, in
addition to the convolved King-profiles, Fig.~\ref{ucdbrightcog} shows three convolved
Sersic profiles with Sersic parameter $n=2$. The curve
corresponding to $r_{\rm eff}=2$ pixel (0.1$''$) fits best the observed
surface brightness of the two
bright UCD candidates, yielding $r_{\rm eff}\simeq$ 310 pc at the
distance of Abell 1689.
This finding supports the
hypothesis that these objects are dwarf galaxies {\it in the process of disruption}
and therefore still possess to some degree a stellar envelope.
Their spectroscopic confirmation as cluster members 
would prove the stripping hypothesis. A possible explanation for the still ongoing
stripping process might be that Abell 1689 consists of at least two sub-clusters
separated in radial velocity (Teague et al.\cite{Teague90}, Girardi et
al.\cite{Girard97}). This indicates that a merger process may be going on,
possibly feeding the center of the cluster with ``fresh'' dwarf galaxies to
be stripped.

Note that both the size and luminosity of the two resolved UCD
  candidates are a
factor 2-3 higher than those of M32's bulge (Graham~\cite{Graham02}). It is
therefore also possible that the latter objects are the bulges of stripped
spiral galaxies, which is the favoured origin for M32 
by Graham's~\cite{Graham02} analysis of M32's surface brightess profile. 
This would imply the existence of a high surface
brightness bulge -- fit well by a Sersic-profile with $n<4$ -- 
and possibly an extended low surface brightness exponential disk. Given that the two UCD
candidates are only marginally resolved, this is extremely difficult to check. From
Fig.~\ref{ucdbrightcog}
it is clear that a single Sersic profile provides a good fit to the surface
brightness profile of the two resolved UCDs.
It is not necessary to include another component into the fit. Nevertheless,
we cannot reject the possibility that the bright UCD candidates are bulges
of stripped spirals, maybe with the outer exponential disk destroyed
  almost completely.

\subsection{Stellar super-clusters?}

\noindent The subclustering of Abell 1689 in radial velocity as mentioned in
the previous section also opens up the possibility for an additional source of bright stellar
clusters, outlined by Fellhauer \& Kroupa~\cite{Fellha02}. They show that in a
merger process very luminous stellar ``super-clusters'' can be created, which
after aging 10 Gyrs resemble the properties of UCDs in Fornax. One existing
example is W3 (Maraston et al.~\cite{Marast04}), an extremely
bright young super-cluster in NGC 7252 of age about 300 Myr, which has
$M_V\simeq -$ 16 mag. In the case of
Abell 1689, the very bright UCD candidates may be very young super-clusters.
However, their color should be much
bluer than found here for Abell 1689. The color range of $1.3<(g-i)<1.8$ for
the UCD candidates with $i<25$ mag corresponds roughly to a restframe color
of $1.0<(V-I)<1.30$
mag, see Fig.~\ref{cmdall}. Using Worthey~\cite{Worthe94}, the expected color for a stellar
population of age 300 Myr is about $(V-I)=0.6$ $\pm$ 0.1 mag,
depending on metallicity. Therefore, the very bright UCD candidates cannot be
explained by the ``super-cluster'' scenario. There are, however, 3 more
unresolved sources at bluer colors than our UCD selection window, see
Fig.~\ref{cmdall}. They have $i\simeq 23.5$ ($M_V\simeq -15.5$) mag and
$(g-i)\simeq 0.6$ mag, which is of the order of what would be expected for
stellar super-clusters. However, for two of these three, the photometric
redshift is 0.01, consistent with them being either foreground stars or
intergalactic globular clusters. For the third object (the reddest one), the
photometric redshift is 0.25 $\pm$ 0.08, consistent with that of the cluster.
See Fig.~\ref{images} for a thumbnail.

Hence, we might have found one analog to W3 in Abell 1689 at a bluer color
than the UCD candidates. Spectroscopic measurements are necessary to prove
this assumption.

\subsection{Foreground stars?}

\noindent Another possibility for the origin of the very bright UCD
candidates is that some of them are foreground stars located in the Milky Way
or its halo, especially the four ones which are definitively unresolved. The
color range of the very bright UCD candidates of $1.3<(V-I)<1.55$ corresponds
to K and M type subdwarfs (Gizis \& Reid~\cite{Gizis99}). The absolute
brightness of these subdwarfs is about $M_V\simeq 9$ mag, corresponding to a
distance of about 10 kpc if the very bright UCD candidates actually are
subdwarfs. The very bright UCD candidates extend about 3 mag in luminosity, or
a factor of about 4 in distance if interpreted as subdwarfs with identical
absolute luminosity. This means they would occupy the distance range 5-20 kpc.
The field of view of the ACS (3 $\times$ 3$'$) corresponds to about 9 $\times$
9 pc at 10 kpc distance. This translates into a volume of about 10$^6$ pc$^3$.
The space density of halo stars at a distance of 10 kpc is about 10$^{-6}
pc^{-3}$ (Kerber et al.~\cite{Kerber01}, Phleps et al.~\cite{Phleps00}). This
shows that some of the very bright UCD candidates could be halo subdwarfs.

However, note that the
mean photometric redshift of the six very bright UCD candidates is 0.245 $\pm$
0.06, consistent with that of the cluster. These photometric redshifts, as all
the ones quoted in this paper, were calculated using the Bayesian BPZ code
by  Benitez~\cite{Benite00}. This involves the assignment of redshift dependent
prior probabilities to six different galaxy templates ranging from an elliptical to an
irregular galaxy SED. 
It is clear that for the six brightest UCD candidates, a statistical approach
like the Bayesian one
might not be
very reliable any more. Some of the template SEDs used in the BPZ algorithm 
-- probably the late-type ones -- may not match very well the
expected SED of an UCD. Therefore,
we calculate the photometric redshift of the six very bright UCD candidates
in two additional ways, using the BPZ code by Benitez~\cite{Benite00}:

First, only with stellar templates of K and M type subdwarfs from
Pickles~\cite{Pickle98} without prior probabilities. This is done to
see how many UCD candidates have colors consistent with foreground stars.
Second, with the elliptical template from Benitez et al.~\cite{Benite04},
also without prior probabilities. UCDs are supposed to be early type
stellar populations, and hence an elliptical galaxy template should
be a good approximation of a real UCD spectrum. Note that up to now,
no precise age/metallicity or line-index measurements of UCDs have been 
published.

The results of the calculations are
shown in Table~\ref{zphotcalc}, indicating the photometric redshifts,
their errors and the $\chi ^2$ of the redshift determination. The results
can be summarized as follows:\\
The two slightly resolved candidates 1 and 3 are the only ones which are
inconsistent with stars at $z_{\rm phot}=0$, and match the redshift $z$=0.183 of
Abell 1689 from $z_{\rm phot,ell}$ and $z_{\rm phot,all}$. 
Candidate 2 is more consistent with a star at $z_{\rm
  phot}=0$
than with an early type stellar population in
Abell 1689.
Candidate 4 is
marginally inconsistent with Abell 1689 from $z_{\rm
  phot,all}$,
and more likely to be an early type stellar population at lower redshift, possibly an
intergalactic globular cluster. Comparing the $\chi ^2$ values, 
it is less likely to be a foreground star.
Candidate 5 is consistent with a star at $z_{\rm
  phot}=0$, while both $z_{\rm phot,ell}$ and $z_{\rm phot,all}$ attribute
it to Abell 1689's redshift. $\chi ^2$ is lower for the stellar template,
and hence this candidate has an ambiguous
redshift assignment with slight preference to a foreground star.
Candidate 6's $z_{\rm phot,all}$ is very high, at the upper limit of what is
adopted in this paper as possible photometric redshift for Abell 1689
members. At the same time, it is consistent with a star at $z_{\rm
  phot}=0$ and an early type population at low redshift, with slightly
lower $\chi ^2$ for the star possibility. It seems that candidate 6 is
therefore either a foreground star or intergalactic globular cluster.

Summarizing, UCD candidates 1 and 3 are the least probable foreground stars
and most probable cluster members
out of the six candidates,
due to their resolved morphology and their photometric redshifts. For
candidates 2, 4, 5 and 6, the photometric redshift information is more
ambiguous and does not allow a clear assignment. Among them, candidate 6 is the 
least probable cluster member. 

It is therefore concluded that there are up to
 4 foregound stars among the 6 brightest UCD candidates, with the
corresponding number of Abell 1689 members being between 5 and 2.
Spectrosopic membership confirmation is needed for more specific statements.

\subsection{Back- or foreground galaxies?}
\noindent Another clear possibility for the identity of the brightest UCD candidates
is that they are galaxies / globular clusters which are not members of Abell 1689. The FOV
towards Abell 1689 in the investigated color range is dominated by cluster
members, but as we consider only six objects, statistical
fluctuations might cause a substantial fraction of these six objects to be
in front or behind the cluster. With the photometric redshifts listed in
table~\ref{zphotcalc}, it is clear that candidates 2, 4 and 6 have either
$z_{\rm phot,ell}$ or $z_{\rm phot,all}$ inconsistent with the redshift of
Abell 1689. Among these, candidate 4 is most likely to be a foreground
object, most possibly an intergalactic globular cluster. Candidate 2 could
be a background galaxy, but note that its $z_{\rm phot,all}$ is less than 2 $\sigma$
away
from the redshift of Abell 1689. Candidate 6 has very low $z_{\rm
  phot,ell}$ and very high $z_{\rm phot,all}$ at comparable ${\chi}^2$. For
this
object, the question whether its colors are created by intrinsically red
populations
at low redshift or blue populations at higher redshift cannot be answered
with certainty.

In conclusion, among the six very bright UCD candidates there are 3
 probable
or possible non-cluster and non-stellar objects:
one probable foreground globular cluster, one possible background galaxy and one
probable non-cluster member, which can be either fore- or background.

\section{Summary and conclusions}
\label{conclusions}
In this paper, the distribution in color, magnitude and space of Ultra
Compact Dwarf (UCD) galaxy candidates (Mieske et al.~\cite{Mieske04},
Drinkwater et al.~\cite{Drinkw03}) in the central 92$''$ (285 kpc) of
Abell 1689 (z=0.183, $(m-M)=39.74$ mag) has been investigated.
The UCD candidates were selected from deep ACS images in g,r,i,z, based on
their magnitude ($i<28$ or $M_V<-11.5$ mag), size (unresolved) and
color. The color windows were $1.2 < (g-i) < 1.8$ mag and $0.38<(r-z)<0.73$
mag, defined with the Bayesian photometric redshift code BPZ by
Benitez~\cite{Benite00}. There are 160 UCD
candidates in Abell 1689 with
$22<i<28$ mag. Combining photometric redshifts
and the radial
density distribution of the UCD candidates shows that about 100 of the 160
UCD candidates are cluster members. In the investigation of
their properties, the following results are obtained:
\begin{enumerate}

\item
The UCD candidates extend to
$M_V\simeq -17.5$ mag, about 4 mag brighter than in the Fornax cluster.
Their luminosity distribution for $26.8<i<28$ ($-12.7<M_V<-11.5$) mag 
is approximated well by the bright end of a Gaussian globular cluster luminosity function
(GCLF) shifted to Abell 1689's distance, implying a total number of about
3-10$\times 10 ^4$ GCs.
 For $i<26.8$ mag, the UCD candidates define an
overpopulation
with respect to the GCLF of about a factor of 4, as ~20 objects are
discovered when about 5 are predicted. This overpopulation is of the order
of the number of UCDs expected from the threshing scenario (Bekki et
al.~\cite{Bekki03}) in Abell 1689. 

\item
The UCD candidates follow a color-magnitude trend with a slope similar to
that defined by the genuine dwarf galaxies in Abell 1689, but shifted
somewhat redder. The shift between the two relations corresponds to a 
magnitude difference of 2-3 mag, about 1-2 mag less than what is expected for the
difference
between parent dwarf galaxy and stripped nucleus from the simulations of
Bekki et al.~\cite{Bekki03}. 

\item
The radial distribution of UCD candidates with $i>27$ mag is consistent with that of the GC
system. For $i>26$ mag it is shallower and more consistent with that of the
genuine dwarf galaxy population in Abell 1689.

\item
Two of the three brightest UCD candidates with $M_V\simeq -17$ mag are
slightly resolved on the ACS images, with implied King-profile core radii of
~35 pc and effective radii of about 300 pc 
at the distance of Abell 1689. These sizes and luminosities are about 2-3 times
higher than the values found for M32's bulge by Graham~\cite{Graham02}.

\item 
Photometric redshifts obtained with late type stellar templates and an elliptical
galaxy
template support the assignment of the two resolved UCD candidates to Abell
1689 based on the original Bayesian photometric redshift calculation.
However, they also allow for up to 4 foreground stars among the six brightest UCD candidates.

\item
There are three stellar super-cluster candidates (Fellhauer \& Kroupa
\cite{Fellha02}) with $M_V\simeq -16$ mag, substantially bluer than the UCD
candidates.

\end{enumerate}

All our findings are consistent with the threshing scenario (Bekki et
al.~\cite{Bekki03}) as a source of at least 10 UCDs for
$i<26.8$ ($M_V<-12.7$) mag. {\it Ultra Compact Dwarf galaxies created by
stripping ``normal'' dwarf or spiral galaxies appear to exist in
Abell 1689.} For $i>26.8$, the globular cluster population of Abell 1689
clearly dominates over possible UCDs created by threshing.
For the following reasons, it appears likely that in the case of Abell 1689 the
threshing
process has not yet finished: (1) the UCD candidates extend to about 4 mag brighter than in
Fornax; (2) their colors are closer to those of
genuine dEs than in Fornax; (3) two of the three brightest UCD candidates are
resolved, implying sizes several times larger than for the UCDs in Fornax.

The next step is clear: spectroscopic confirmation of the cluster membership
assignment 
obtained with photometric redshifts for the UCD candidates in this paper. 
For the very brightest UCD candidates
with $22<i<24$ mag, this should be a
feasible task with an 8m-class telescope. The spectroscopic confirmation
would prove the existence of Ultra Compact Dwarf galaxies at almost 50 times
the distance of the Fornax cluster, the location of the UCD's orginal discovery.
It would imply that the disruption of dwarf or spiral galaxies in the cluster
tides may be a common process. Detailled numerical simulations in order to test
the magnitude of this effect for fainter luminosities are needed. This would
allow to check in how far the tidal disruption might be responsible for the
``missing satellite'' problem, i.e. the disagreement between predicted and
observed
frequency of dark matter halos.
\acknowledgments

\noindent ACS was developed under NASA contract NAS 5-32864, and this research 
has been partially supported by NASA grant NAG5-7697 and by an equipment 
grant from Sun Microsystems, Inc. The Space Telescope Science Institute 
is operated by AURA Inc., under NASA contract NAS5-26555. We are 
grateful to K. Anderson, J. McCann, S. Busching, A. Framarini,
S. Barkhouser, and T. Allen for their invaluable contributions to the 
ACS project at JHU. SM was supported by DAAD Ph.D. grant Kennziffer D/01/35298 and
DFG Projekt Nr. HI 855/1-1. LI would like to acknowledge support from
"proyecto Fondap \# 15010003".

\newpage

\clearpage

\begin{table}
\begin{tabular}{l|l||l|l|l}
&$i$ [mag]&$z_{\rm phot,star}$  (${\chi}^2$) &$z_{\rm phot,ell}$ ($\chi ^2$)
&$z_{\rm phot,all}$ ($\chi ^2$) \\\hline\hline
Cand1&22.18 & 0.091 $\pm$ 0.073 (7.9) & 0.164 $\pm$ 0.078  (3.9) & 0.160$\pm$ 0.076 (7.1) \\
Cand2&22.38 & 0.001 $\pm$ 0.067 (2.6) & 0.188 $\pm$ 0.079  (16.1) & 0.300 $\pm$ 0.076 (18.9)\\
Cand3&22.77 & 0.102 $\pm$ 0.029 (8.8) & 0.180 $\pm$ 0.079  (5.0)& 0.180 $\pm$ 0.078 (5.0)\\
Cand4&23.24 & 0.001 $\pm$ 0.067 (47.2) & 0.001 $\pm$ 0.067  (18.35) & 0.090 $\pm$ 0.072 (12.8)\\
Cand5&23.80 & 0.001 $\pm$ 0.069 (3.3) & 0.142 $\pm$ 0.076  (13) & 0.250 $\pm$ 0.085 (18)\\
Cand6&24.62 & 0.001 $\pm$ 0.067 (24.7) & 0.001 $\pm$ 0.067  (35) & 0.490 $\pm$ 0.098 (36)\\
\end{tabular}
\caption[]{\label{zphotcalc}Photometric redshifts $z_{\rm
    phot}$ for the six brightest UCD candidates, obtained in three different
  ways: $z_{\rm phot,star}$ from using stellar templates of K and M type
  subdwarfs
without prior probabilities. $z_{\rm phot,ell}$ from using the elliptical
galaxy template of Benitez~\cite{Benite04} without priors, too. Finally,
$z_{\rm phot,all}$ from using the entire set of six different galaxy type
  templates
with prior probabilities (see Coe et al.~\cite{Coe04} and Broadhurst
  et al.~\cite{Broadh04}). Errors are 1 $\sigma$. The $\chi ^2$ values 
show how well template and object colors match at the
  calculated photometric redshift.}
\end{table}
\end{document}